\documentclass[12pt,numberedappendix,twocolappendix]{emulateapj}
\usepackage{graphicx}
\usepackage{amsmath}
\usepackage{colortbl}
\usepackage{comment}

\usepackage{natbib,aas_macros}
\newcommand{\greeceIII}{\text{I\hspace{-.1em}I\hspace{-.1em}I} }
\newcommand{\greeceII}{\text{I\hspace{-.1em}I}}

\usepackage{hyperref}   
\hypersetup{colorlinks=true,linkcolor=blue,citecolor=blue,filecolor=blue,urlcolor=blue}

\begin{document}
\title{Cosmological Simulations of Early Black Hole Formation:
Halo Mergers, Tidal Disruption, and the Conditions for Direct Collapse}

\setcounter{footnote}{1}
\author{Sunmyon Chon\altaffilmark{1}} 
\author{Shingo Hirano\altaffilmark{1}}
\author{Takashi Hosokawa\altaffilmark{1,2}}
\author{Naoki Yoshida\altaffilmark{1,3}}
\altaffiltext{1}{Department of Physics, School of Science,
  University of Tokyo, Bunkyo, Tokyo 113-0033, Japan}
\altaffiltext{2}{Research Center for the Early Universe,
  University of Tokyo, Bunkyo, Tokyo 113-0033, Japan}
\altaffiltext{3}{Kavli Institute for the Physics and
  Mathematics of the Universe (WPI), Todai Institutes for Advanced Study,
  University of Tokyo, Kashiwa, Chiba 277-8583, Japan}

\begin{abstract}
\textcolor{black}{
  Gravitational collapse of a massive primordial gas cloud is thought to be
  a promising path for the formation of 
  supermassive black holes in the early universe.
  We study conditions for the so-called direct collapse (DC) black hole formation
  in a fully cosmological context. 
  We combine a semianalytic model
  of early galaxy formation with halo merger trees constructed
  from dark matter $N$-body simulations. We locate a total of
  68 possible DC sites in a volume of
  $20\;h^{-1}\;\mathrm{Mpc}$ on a side. 
  We then perform hydrodynamics simulations for 42 selected halos
  to study in detail the evolution of the massive clouds within them.
  We find only two successful cases where the gas clouds rapidly
  collapse to form stars. 
  In the other cases, gravitational collapse is prevented by
  the tidal force exerted by a nearby massive halo, which
  otherwise should serve as a radiation source necessary for DC.
  Ram pressure stripping disturbs the cloud approaching the source.
  In many cases, a DC halo and its nearby
  light source halo merge before the onset of cloud collapse.
  Only when the DC halo is assembled through major mergers, 
  the gas density increases rapidly to trigger gravitational instability. 
  Based on our cosmological simulations, we conclude that the event
  rate of DC is an order of magnitude smaller than reported
  in previous studies, although the absolute rate is still poorly
  constrained. It is necessary to follow the dynamical evolution
  of a DC cloud and its nearby halo(s) in order to determine
  the critical radiation flux for DC.}
\end{abstract}

\section{Introduction}
The existence of luminous quasars at $z > 6$
suggests rapid formation and growth of supermassive black holes (SMBHs) 
in the early universe \citep[e.g.,][]{Mortlock+2011,Wu+2015}. 
Theoretically, very efficient mass accretion is needed
for the massive BHs to be in place when the age of the Universe
was less than $10^9~$yr.
Even with the maximum Eddington accretion rate,  
a $100 \;M_\odot$ seed BH can marginally achieve
mass growth to $\sim 10^9\;M_\odot$ by $z \simeq 7$.

The largest mass of the stellar-mass BHs
is $20 - 30 \;M_\odot$ in the local universe
\citep{Casares+2014,Belczynski+2010},
but Population III (Pop~III) remnant BHs might have larger masses of 
$\sim 100 \;M_\odot$. In fact, recent theoretical studies suggest
that Pop III stars have 
a broad mass range extending beyond a few hundred solar masses
\citep[e.g.,][]{Hirano+2014,Hirano+2015,Susa+2014}. 
Stellar evolution calculations predict that, 
with small mass-loss rates, Pop III stars heavier than 
$300\;M_\odot$ will leave BHs with essentially the same mass \citep{HegerWoosley2002}. 
Although such massive BHs might appear promising SMBH seeds,
maintaining the Eddington accretion over nearly
$10^9~$yr is difficult, if not unphysical,
because a variety of radiative feedback effects from stars and accreting BHs reduce the
gas accretion rate considerably \citep[e.g.,][]{Yoshida2006, Milos+2009, PR2011}.

The so-called direct collapse (DC) model posits
formation of $10^5 - 10^6 \;M_\odot$
supermassive stars (SMSs) in the early universe.
Such a SMS leaves, at the end of its evolution, 
an equally massive BH (hereafter referred to as DCBH), 
which can kickstart the rapid mass evolution to SMBHs 
\citep[e.g.,][]{Volonteri2010, DiMatteo+2012, Valiante+2016}.
An important question, then, is how often or how rarely
DC occurs in the early universe to connect with the observation. 
The DC model actually assumes multistep
processes under a few critical conditions. 
First, the gas in an atomic-cooling halo
collapses nearly isothermally at
$T \simeq 8000~{\rm K}$ if the formation of hydrogen molecules
is suppressed \citep{Omukai2001}.
Three-dimensional simulations show no signatures
of vigorous fragmentation in the early collapse stage 
\citep[e.g.,][]{BL03, Inayoshi+2014, Becerra+2015}, and thus
a single embryo protostar is formed at the cloud's center. 
The protostar then accretes the surrounding gas at a very large rate of  
$\dot{M} \sim 0.1 - 1\;M_\odot~ \mathrm{yr}^{-1}$.
One can naively expect the well-known relation for a contracting gas cloud,
$\dot{M} \sim M_\text{J}/ t_\text{ff} \propto T^{3/2}$, 
where $M_\text{J}$ is the Jeans mass and $t_\text{ff}$ 
is the free-fall time. Because the temperature $T\simeq 8000$~K is much higher
than in the ordinary Pop III formation case with $T\simeq 200-1000\;\mathrm{K}$,
the gas mass accretion rate can be large in the DC model.
In fact, simulations following the evolution in the late accretion
phase demonstrate that such rapid accretion is realized
\citep[e.g.,][]{Latif+2013}.
Since the lifetime of a very massive star is typically a few million years, 
the final stellar mass can reach $\sim 10^5 - 10^6 \;M_\odot$.
Hence, this is clearly an appealing model if the above physical processes
take place in order.

A variety of models have been proposed to set environments free from 
$\mathrm{H}_2$ formation and cooling \citep[e.g.,][]{BL03, Inayoshi+2012, Tanaka+2013}. 
$\mathrm{H}_2$ dissociation by far-ultraviolet 
(FUV) radiation is one of the major processes.
The intensity required for DC, often denoted by $J_{21}^\text{crit}$,
\footnote{$J_{21}$ is the intensity at the Lyman-Werner bands 
normalized in the units of 
$10^{-21}\;\mathrm{erg\;s^{-1}\;cm^{-2}\;Hz^{-1}\;str^{-1}}$. }
has been derived by several authors 
\citep[e.g.,][]{Omukai2001, Shang+2010, Latif+2014, Sugimura+2014, Regan+2014}.

There are actually two dissociation processes: 
one is photodissociation of $\mathrm{H}_2$ molecules
\citep[]{StecherWilliams1967}, 
\begin{equation}
\mathrm{H}_2 + \gamma \longrightarrow 2\mathrm{H},                \label{pd_H2}
\end{equation}
and the other is photodetachment of $\mathrm{H}^-$,
the catalyzer of the $\mathrm{H}_2$ formation via 
\begin{equation}
\mathrm{H}^-  + \gamma \longrightarrow \mathrm{H} + \mathrm{e}^- . \label{pd_Hminus}
\end{equation}
Reaction \eqref{pd_H2} is only induced by FUV photons  
with energy $11.2\;\mathrm{eV} < h\nu < 13.6\;\mathrm{eV}$ in the
Lyman-Werner (LW) bands,
and reaction \eqref{pd_Hminus} has a finite cross section 
for lower-energy photons with $h \nu > 0.58 \;\mathrm{eV}$. 
Thus, the critical intensity
$J_{21}^\text{crit}$ depends on the exact spectrum
of the incident LW radiation, i.e., the nature of the radiation
source. For example,
detailed 3D simulations of \cite{Shang+2010} show that 
\begin{eqnarray}
\label{eq:jcr_pop}
J_{21}^\text{crit} =\left\{ \begin{array}{ll}
10^4 & \text{(for Pop III sources) ,}\\
100 & \text{(for Pop II sources) ,}\\
\end{array} 
\right .
\end{eqnarray} 
where the black body spectra of $T_{\rm eff} =10^5 \;\mathrm{K}$ and $10^4
\;\mathrm{K}$ are assumed for the Pop III and II sources, respectively.

Considering necessary conditions for DC such as 
$J_{21}^\text{crit}$, many authors have calculated
the cosmological event rate. 
\cite{Dijkstra+2008, Dijkstra+2014} use a
Monte Carlo model to count the number of halos irradiated by strong
LW radiation exceeding $J_{21}^\text{crit}$. 
In these studies, relevant baryonic physics are implemented 
in a phenomenological manner to search DC candidates in a large
spatial volume ($\sim$~Gpc) considering its parameter dependencies.
They study how the baryonic processes, such as
how far the metals are dispersed from galaxies,
affect the number density of DC.
\cite{Agarwal+2012} 
employ a semianalytic model of star and 
galaxy formation with halo merger trees constructed from
cosmological $N$-body simulations.
\cite{Johnson+2013} perform cosmological
smoothed particle hydrodynamics (SPH)
simulations to directly follow the evolution 
of DC candidate halos and the gas clouds within them.
\cite{Habouzit+2016} use adaptive mesh refinement
simulations to follow 
hydrodynamical evolution of several DC candidate halos
and the gas clouds within them with various feedback models
and $J_{21}^\text{crit}$. 
The DC event rate estimates in these studies appear to be
very large, in the light of the observed quasar abundance at $z>6$, 
$\text{a few} \;\mathrm{Gpc}^{-3}$.
More importantly, many of the previous studies do not follow
the DC cloud {\it collapse} and thus cannot determine robustly whether
or not SMSs are formed in the DC halos.

In this paper, we critically examine whether DC is triggered
in early dark matter (DM) halos by using direct hydrodynamics simulations.
We first locate DC candidate halos in a large cosmological
volume by combining a semianalytic model and halo merger trees.
We then follow the dynamical evolution of the selected 
$42$ halos to examine explicitly whether the gas clouds
collapse gravitationally. 
We show that the gas density does not increase in 
most of the candidate halos, largely owing to strong tidal
force caused by nearby massive halos. 
We find only two successful cases
where the collapse is accelerated by halo mergers
despite the strong tidal force. 
The two cases clearly present a viable path for DC in a fully
cosmological context. Based on our simulations,
we discuss the necessary conditions for DC.

The organization of the paper is as follows.
In Section 2, we describe the numerical method. 
In Section 3, we present 
the statistical properties of the candidate halos.
Section 4 shows that the hydrodynamical evolution of gas clouds 
in the candidate halos for representative cases.
The cloud collapse advances until the density
reaches $10^8\;\mathrm{cm^{-3}}$ for two cases, 
which is described in Section 5.
We finally discuss implications in Section 6,
and give concluding remarks in Section 7.

Throughout this paper, we use the cosmological parameters provided by 
the Planck Collaboration results \citep{PlanckXVI2014}, which are 
$\Omega_\text{m} = 0.308$,
$\Omega_\text{b} = 0.0483$, $\Omega_\Lambda = 0.692$, normalized 
Hubble constant $h = 0.677$, and $\sigma_8 = 0.8288$. 

\section{Methodology}

We perform a set of multiscale numerical simulations
to study the DC model in a realistic cosmological context. 
We first carry out cosmological $N$-body simulations to 
construct halo merger trees.
We develop a semianalytic model of star and galaxy formation 
and implement it for the halo merger trees.
In our model, metal enrichment and buildup of LW radiation field 
are followed in order to locate potential sites for DC.
We adopt the conditions for  DC 
that are widely used in the literature (see Section~\ref{ssec:criteria} below).
For about 40 DC sites selected in this manner, 
we perform zoom-in $N$-body/SPH
simulations to follow the thermal evolution and
gravitational collapse of the gas clouds.

\subsection{Parent $N$-body Simulations} \label{sec_parentNbody}

We use the parallel Tree-PM code
Gadget-2 \citep{Springel2005} to run our parent
cosmological simulation.
The initial condition is generated at $z_\text{ini} = 99$ 
by MUSIC \citep{HahnAble2013}, which is based on second-order 
Lagrangian perturbation theory. The box size is 
$20\;h^{-1}\;\mathrm{Mpc}$ on a side. 
The number of DM particles is $256^3$ and the particle mass is $4.08\times
10^7\;h^{-1}\;M_\odot$. The simulation is stopped at $z = 9$.
Halos are identified by the friend-of-friend (FOF) algorithm with
linking parameter $b = 0.2$.

Zoom-in simulations are performed to study 
the detailed evolution and structure of the regions that
contain potential DC sites.
To this end, we first mark 10  most massive halos in our
parent simulation, which correspond roughly to $3$--$4\sigma$
density peaks.
We then regenerate the initial conditions for
the zoom-in regions with increasing mass resolution
and rerun $N$-body simulations. 
The volume of a zoomed-in region is
$2.0\;(h^{-1}\;\mathrm{Mpc})^3$, and the refined particle mass is 
$1.2\times 10^3\;h^{-1}\;M_\odot$.

In the zoomed-in region,  minihalos with 
$M_\text{halo} \sim 10^5\;M_\odot$, possible Pop III
star hosts, are resolved by more than 100 DM particles. 
This is sufficient to follow Pop III star formation 
and the associated metal enrichment of the surrounding gas.

\subsection{Construction of Merger Tree}

We utilize the outputs of the zoom-in simulations
to construct halo merger trees.
The time interval of the two adjacent outputs is $10\;\mathrm{Myr}$,
which is shorter than the dynamical time of collapsing halos at $z < 40$. 
With the frequently dumped snapshots, we can
trace the formation history of Pop III stars.

Merger trees are constructed by tracing the halo member particles.
Progenitors and descendants are determined for each halo. A simple
particle tracking method often fails because
a halo can split into multiple descendants. Therefore, we employ the
subhalo-based merger tree algorithm \citep{Springel+2005Natur}. 
Hereafter, we use the term ``halo'' also for a subhalo.

Subhalos in each FOF halo are defined by SUBFIND
\citep{Springel+2005Natur}. We trace the subhalo member particles 
to find a descendent halo(s).
This method assigns a unique descendant halo for 99\% objects,
while the remaining 1\% have still multiple descendants. 
The multiple descendants appear because
our time resolution is indeed very short and 
transient structure of halo mergers can be detected. 
We simply regard the multiple descendants found in the exceptional
cases as a single object.

We use the following five major halo properties: 
the halo mass $M$, formation redshift $z$,
the virial temperature $T_\text{vir}$, 
viral radius $R_\text{vir}$, and circular velocity $V_\text{c}$,
which are defined as follows:
\begin{eqnarray}
T_\text{vir} &=& 1.98 \times 10^4 \left ( \frac{\mu}{0.6} \right ) \left [ \frac{\Omega_\text{m}}{\Omega_\text{m}(z)} \frac{\Delta_\text{c}}{18\pi^2} \right ]^{1/3} \nonumber \\
&&\;\;\;\; \left ( \frac{M}{10^8\;h^{-1}\;M_\odot} \right )^{2/3} \left ( \frac{1+z}{10} \right ) \;\;\;\; \mathrm{K}, \label{eq_Tvir} \\
R_\text{vir} &=& 0.784 \left [ \frac{\Omega_\text{m}}{\Omega_\text{m}(z)} \frac{\Delta_\text{c}}{18\pi^2} \right ]^{-1/3}  \nonumber \\ 
&&\;\;\;\; \left ( \frac{M}{10^8\;h^{-1}\;M_\odot} \right ) ^{1/3}  \left ( \frac{1+z}{10} \right )^{-1} h^{-1}\;\mathrm{kpc}, \\
V_\text{c} &=& 23.4 \left [ \frac{\Omega_\text{m}}{\Omega_\text{m}(z)} \frac{\Delta_\text{c}}{18\pi^2} \right ]^{1/6}  \nonumber \\
&&\;\;\;\; \left ( \frac{M}{10^8\;h^{-1}\;M_\odot} \right ) ^{1/3}  \left ( \frac{1+z}{10} \right )^{1/2} \mathrm{km\;s^{-1}},
\end{eqnarray} 
where $\mu$ is the mean molecular weight,
$\Omega_\text{m}(z)$ is the fraction of the energy budget of the matter at $z$,
and $\Delta_\text{c}$ is the critical overdensity for the halo usually set to
be $18\pi^2$ \citep{Barkana+2001}.

\subsection{Semianalytic Model for Star and Galaxy Formation} 
\label{sec_SAM}

We employ a semianalytic model of
star formation and galaxy formation.
We largely follow the method in \cite{Agarwal+2012} and briefly
describe the implemented processes in the following subsections.

\subsubsection{Pop III Star Formation and Metal Enrichment} \label{popIII formation}

Pop III star formation is initiated by the onset of $\mathrm{H}_2$
cooling, which becomes efficient in halos with 
$T_{\rm vir} \gtrsim 2000\;\mathrm{K}$ \citep[e.g.,][]{Tegmark+1997, Yoshida+2003}.
We assume that a Pop III star forms in a halo when the
virial temperature exceeds $2000\;\mathrm{K}$.
We adopt a fixed stellar mass of $100\;M_\odot$ for the
Pop III stars \citep[e.g.,][]{Hirano+2015}.
Our conclusions are not significantly affected by the choice
of the Pop III stellar mass (see Section \ref{sec_SF_LW}).
The lifetime of a Pop III star is set to be 2~Myr \citep{Schaerer2002}. 
We assume that all the gas in the host halo 
is promptly polluted by heavy elements dispersed by a supernova (SN) explosion.
Hereafter we refer to the halos before and after the Pop III star formation
as ``pristine'' and ``metal-enriched'', respectively.

\subsubsection{Pop II Star Formation} \label{SAM}

After a massive Pop III star dies as an SN, 
further star formation will be suppressed in the same halo
for a while because
the gas is evacuated by the energetic SN. 
When the halo grows sufficiently to attain a mass exceeding a threshold
value $M_\text{crit}$, continuous Pop II
star formation begins. The threshold mass is given essentially
by the condition for efficient gas cooling.
We fix $M_\text{crit}$ to be $10^7\; h^{-1}\;M_\odot$ \citep{LF2012}.
We label the halos with ongoing Pop II star formation
as ``star-forming''.

We adopt a three-component model for
the baryons; ``hot gas'', ``cold gas'', and ``stars''
coexist in a halo.
We model the evolution of the baryonic components
as follows.

\begin{enumerate}
\setlength{\itemsep}{-0.5mm}
\setlength{\leftskip}{5.0mm}

\item Halo growth: 
When a halo mass increases by $\Delta M$,
an accompanying gas of $\Delta M_{\rm gas} = f_\text{b} \Delta M$ is
added and it is shock-heated
to the halo virial temperature, where $f_\text{b} \equiv \Omega_\text{b}/\Omega_\text{m}$.

\item Cooling:
The hot gas component then cools radiatively. 
We assume that the cooling timescale is approximately
given by the local dynamical time 
$t_\text{dyn} \equiv R_\text{vir}/V_\text{c}$.

\item Star formation: 
The ``cold gas'' component is then transformed into stars
over the characteristic timescale of  
$t_\text{SF} \equiv 0.1t_\text{dyn}/\alpha$
\citep[e.g.,][]{Kauffmann+1993}, where $\alpha$ is the star formation
efficiency, which is the mass fraction of the cold gas converted 
into the stars. We set $\alpha=0.005$.

\item Feedback:
An SN explosion returns a part of the
cold gas component into hot gas by energy
injection. We adopt the mass-converting rate $\dot{m}_\text{reheat}$ that is
proportional to the star formation rate (SFR) $\dot{m}_\text{star}$,
\begin{eqnarray}
\dot{m}_\text{reheat} = \gamma \dot{m}_\text{star}
= \gamma \frac{m_{\rm cold}}{t_{\rm SF}}
, \\
\gamma = \left (\frac{V_\text{c}}{V_\text{out}} \right )^{-\beta}
\end{eqnarray}
\citep{Cole+1994}.
We adopt $\beta=2.0$ and $V_\text{out} = 110\;\mathrm{km\;s^{-1}}$ 
in our calculations \citep{DekelSilk1986}.
\end{enumerate}

The four processes presented above can be described by
the following differential equations:
\begin{eqnarray}
\frac{\mathrm{d}m_\text{hot }}{\mathrm{d}t} &=& -\frac{m_\text{hot}}{t_\text{dyn}} + \gamma \frac{m_\text{cold}}{t_\text{SF}} + f_\text{b} \frac{\mathrm{d}m_\text{halo}}{\mathrm{d}t}  , \nonumber \\
\frac{\mathrm{d}m_\text{cold}}{\mathrm{d}t} &=& \frac{m_\text{hot}}{t_\text{dyn}} - (1+\gamma) \frac{m_\text{cold}}{t_\text{SF}}, \\
\frac{\mathrm{d}m_\text{star}}{\mathrm{d}t} &=& \frac{m_\text{cold}}{t_\text{SF}}.  \nonumber
\end{eqnarray}
We solve these equations explicitly for each time step.

\subsection{LW Radiation Field} \label{sec_LW_field}

The formed stars build up UV background radiation,
which causes significant impact on early structure formation 
through $\mathrm{H}$ ionization and $\text{H}_2$ dissociation. 
The intergalactic medium (IGM) is optically thick for ionizing photons before
the cosmic reionization, and thus the effects of ionizing photons
are relatively limited.
We do not consider ionizing photons in our semianalytical calculations,
but we examine the effect using one-zone calculations in Section \ref{sec_ionization_photons}.

$\text{H}_2$ dissociating photons easily propagate throughout
the IGM and affect star formation in other (distant) halos. 
Photons in the Lyman-Werner bands are emitted from both Pop III and II 
stars. 
The stellar luminosities are calculated as follows:
\begin{enumerate}
\setlength{\itemsep}{-0.5mm}
\setlength{\leftskip}{5.0mm}

\item Pop III radiation: 
We use the model spectrum of the $100\;M_\odot$ star provided by
\cite{Schaerer2002}. 

\item Pop II radiation:
The spectrum of a galaxy is calculated by the population synthesis model
STARBURST99 \citep{Leitherer+1999}. 
We adopt the Kroupa stellar initial mass function (IMF) and
metallicity $Z = 0.001\;Z_\odot$. 
The total luminosity at
the LW bands can be obtained by time-integrating the stellar luminosities
over the star formation history, 
\begin{eqnarray}
L_{\text{LW},0}(t) &=& \int _{11.2\;h^{-1}\mathrm{eV}}^{13.6\;h^{-1}\mathrm{eV}} L_{\nu, 0}(t) \mathrm{d}\nu, \\
L_{\text{LW},\text{tot}}(t) &=& \int _{0}^{t_0} \mathrm{d}t' \; \dot{m}_\text{star}(t - t') L_{\text{LW},0}(t'),
\end{eqnarray}
where $L_{\text{LW},0}(t)$ and $L_{\nu, 0}(t)$ are the
luminosity and the luminosity differentiated per unit wavelength 
radiated by the stars formed at time $t$, respectively,
and $L_{\text{LW},\text{tot}}(t)$ is the total luminosity
including the contribution of all the stars formed in the past.
The reference time $t_0$ is the time when the Pop II star formation begins.
\end{enumerate}

\subsubsection{Pop III Star Formation under the LW Radiation}
The condition of the Pop III star formation under LW radiation is
investigated by \cite{Machacek+2001}, who parameterize the criterion for 
Pop III star formation under the LW radiation as
\begin{equation}
M_\text{th} = \psi \left ( 1.25 \times 10^5 + 2.8 \times 10^6 \;J_\text{21}^{0.47} \right ) M_\odot, 
\end{equation}
where $M_\text{th}$ is the threshold halo mass above
which Pop III stars are formed.
Note that $\psi \simeq 4$ is a correction factor introduced by \cite{ON2008}. 
For halos irradiated by nearby radiation sources, we use the
criterion of the Pop III star formation $M_\text{halo} > M_\text{th}$ 
instead of $T_\text{vir} > 2000\; \mathrm{K}$.

\subsection{Selection of DC Halos}
\label{ssec:criteria}

We impose the following three criteria for DC candidate halos:
\begin{enumerate}
\setlength{\itemsep}{-1.mm}
\item the halo is chemically pristine;
\item it is irradiated by sufficiently strong LW radiation with
$J_{21} > J_{21}^\text{crit}$; and
\item it is massive enough for atomic hydrogen cooling to operate, i.e.,
$T_\text{vir} > 8000\;\mathrm{K}$.
\end{enumerate}
The first criterion is required to avoid efficient metal line and dust cooling. 
The second condition prevents prior Pop III star formation by suppressing $\text{H}_2$ cooling. 
The last criterion ensures rapid cooling of a massive gas cloud. 
Here, $J_\text{crit}$ is set to be $10^4$ and $100$
for Pop~III and Pop~II light sources, respectively, as in eq. \eqref{eq:jcr_pop}.
Note that the assumed spectrum
is different from the spectrum calculated
by the population synthesis method (Section \ref{sec_LW_field}).
We deliberately made this choice. We give further discussion
on $J_\text{crit}$ in the next section.

Other conditions are also conceivable for DC.
For example, the angular momentum of a contracting gas cloud
should not be very large to significantly reduce
the gas mass accretion rate.
Clearly, further direct hydrodynamical simulations following 
the cloud collapse are necessary to assess additional, if any, conditions. 
This is indeed the main reason why we perform costly hydrodynamics
simulations in the present study.

\subsection{Hydrodynamical Simulation} \label{sec_hydro}
The evolution of the gaseous components
in the DC candidate halos is followed using SPH+Tree code
Gadget-3 \citep{Springel2005}.
We include 14 primordial species ($\text{e}^-$,
$\text{H}$, $\text{H}^+$, $\text{He}$, $\text{He}^+$, $\text{He}^{2+}$,
$\text{H}_2$, $\text{H}_2^+$, $\text{H}^-$, $\text{D}$, $\text{D}^+$,
$\text{HD}$, $\text{HD}^+$, and $\text{D}^-$) and follow the chemical
and cooling processes, including
$\text{H}_2$ photodissociation caused by 
external LW radiation from nearby light sources.
The temporally and spatially varying LW intensity is given
by the semianalytic calculations 
described in Section \ref{sec_LW_field}.
We consider attenuation of the LW intensity by gas self-shielding 
using the six-ray evaluation scheme \citep{Yoshida+2003} 
for $\text{H}_2$ and $\text{HD}$ molecules 
\citep{WolcottGreen+2011,WGH2011}. 

The initial DM particle distribution is extracted from the parent
$N$-body simulation (Section \ref{sec_parentNbody}). 
The gas particles are displaced at the half mesh apart from the DM particle,
where the length of one mesh corresponds to the mean separation 
of the DM particles.
The masses of the gas and DM particles are $1.8\times 10^2$ and 
$1.0\times 10^3 ~h^{-1}M_\odot$, respectively.

We follow the cloud collapse in the DC candidate halos until
the halo merges with a nearby metal-enriched halo or until
the central gas density reaches $10^8\;\mathrm{cm}^{-3}$.
During the cloud collapse, we refine particles 
so that the local Jeans
mass is always larger than 1000 times the particle mass
\citep{Truelove1997}.
The practical implementation of our particle
splitting is based on \cite{Kitsionas+2002}.
We have confirmed the convergence of our results with
using a more strict criterion for the particle splitting,
where the Jeans mass is always resolved by more than
$10^4$ particles.

Occasionally, high-density gas clumps in regions far from a
candidate halo cause the simulation time step to be extremely short.
To circumvent this difficulty, we harden the equation of state of the gas 
above threshold densities of $100 - 10^4 \;\mathrm{cm}^{-3}$ for
nontarget halos. We further reduce the computational cost 
by employing a ``de-refinement'' method for nontarget halos,
by which the particles are combined to yield low-resolution particles.
Again, this procedure dramatically reduces the computational time
and enables us to follow a long-term evolution of the candidate halos.
We describe further details in Appendix~\ref{sec_dezoom}. 

In this study, we use the black-body spectrum with $T_\text{eff}=10^4~$K 
for the FUV radiation spectrum.
Our main goal is to study the collapse of the DC halo
starting from the cosmological initial condition and to see the
properties of the cosmologically selected DC halos.
To investigate the effect of the DC halo selection,
we should compare our results with the previous studies
on the same conditions
except with the proper cosmological settings.
Because most of the previous studies 
employ the $T_\text{eff}=10^4~$K black-body spectrum
\citep{Shang+2010,Latif+2014,Latif+2015},
we use the same spectrum here, and thus
we adopt $J_{\rm crit} = 100$ for the Pop II light
sources in Section~\ref{ssec:criteria}.
Note that more realistic spectra of metal-poor galaxies actually 
provide a higher value of $J_{\rm crit} \sim 1000$ \citep{Sugimura+2014}.

\begin{figure}[tp]
	\includegraphics[width=8.cm]{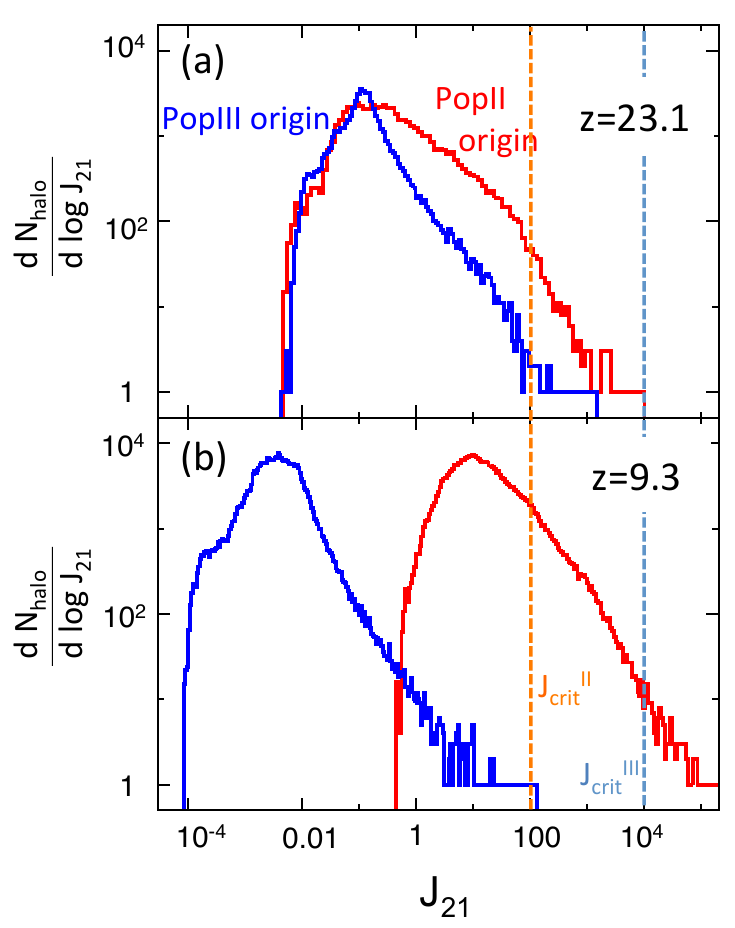}
	\caption{Distributions of the LW intensities $J_{21}^{\greeceIII}$ and
$J_{21}^{\greeceII}$ at the centers of pristine halos 
at (a) $z=23.1$ and (b) $z=9.3$. 
The dashed lines show the critical values for Pop II (red) and Pop III sources (blue). 
	}
	\label{J21_evo}
\end{figure}

\begin{table}[tb]
	\caption[]{properties of DC candidate halos}
\scalebox{0.9}[0.9]{
	\begin{tabular}{l | c  c  c c r  }\hline
	ID	    & $M_\text{halo}$                & z     & $R_\text{vir}$                  & $R_\text{dist}$                & $t_\text{ff}\hspace{2mm}$ \\ 
		    &  $[10^6 h^{-1}\;M_\odot]$ &       & [$h^{-1}\;\mathrm{kpc}$] & [$h^{-1}\;\mathrm{kpc}$] & [Myr] \\ \hline
	DC0 (S1)	&      \hfil $8.37$       	&	21.1		& 0.23 &\hfil $1.27$		&	80 \\ 
	DC1		&	\hfil $6.36$	        &	19.8		& 0.22 &\hfil $0.81$		&	28 \\ 
	DC2 (F2)	&	\hfil $5.91$        	&	19.1		& 0.22 &\hfil $1.79$		&	50 \\ 
	DC3		&	\hfil $6.45$   		&	19.1		& 0.23 &\hfil $1.99$		&	60 \\ 
	DC4		&	\hfil $6.47$		&	19.1		& 0.23 &\hfil $0.99$		&	37 \\ 
	DC5		&	\hfil $6.00$		&	18.7		& 0.23 &\hfil $1.11$		&	50 \\ 
	DC6		&	\hfil $7.32$		&	18.4		& 0.25 &\hfil $2.57$		&	50 \\ 
	DC7		&	\hfil $8.05$		&	17.5		& 0.27 &\hfil $3.93$		&	60 \\ 
	DC8		&	\hfil $6.31$		&	17.5		& 0.25 &\hfil $3.02$		&	35 \\ 
	DC9		&	\hfil $6.21$		&	17.3		& 0.25 &\hfil $4.48$		&	60 \\ 
	DC10	&	\hfil $6.88$		&	17.2		& 0.26 &\hfil $5.24$		&	65 \\ 
	DC11	&	\hfil $7.21$		&	17.2		& 0.27 &\hfil $4.78$		&	120 \\ 
	DC12 	&	\hfil $7.78$		&	16.9		& 0.28 &\hfil $2.79$		&	240 \\ 
	DC13 	&	\hfil $8.12$		&	16.4		& 0.29 &\hfil $0.39$		&	35 \\ 
	DC14 (F1)&	\hfil $5.71$		&	16.4	        & 0.26 &\hfil $3.10$		&	70 \\ 
	DC15	&	\hfil $6.64$		&	16.4   	& 0.27 &\hfil $2.29$		&      85 \\ 
	DC16	&	\hfil $18.3\hspace{2.8mm}$		&	16.4		& 0.38 &\hfil $0.17$		&	65 \\ 
	DC17	&	\hfil $8.69$		&	16.4		& 0.30 &\hfil $3.45$		&	100 \\ 
	DC18	&	\hfil $9.45$		&	16.2		& 0.31 &\hfil $5.85$		&	70 \\ 
	DC19	&	\hfil $7.16$		&	16.0		& 0.28 &\hfil $4.96$		&	75 \\ 
	DC20	&	\hfil $12.3\hspace{2.8mm}$		&	15.9		& 0.34 &\hfil $2.42$		&	25 \\ 
	DC21	&	\hfil $8.07$		&	15.7		& 0.30 &\hfil $5.39$		&	70 \\ 
	DC22	&	\hfil $7.81$		&	15.7		& 0.30 &\hfil $6.88$		&      85 \\ 
	DC23	&	\hfil $7.53$		&	15.5		& 0.30 &\hfil $1.94$		&	50 \\ 
	DC24	&	\hfil $7.92$		&	15.5		& 0.30 &\hfil $1.64$		&	40 \\ 
	DC25	&	\hfil $7.96$		&	15.5		& 0.30 &\hfil $4.06$		&	140 \\ 
	DC26	&	\hfil $8.46$		&	15.3		& 0.31 &\hfil $7.36$		&	120 \\ 
	DC27	&	\hfil $10.0\hspace{2.8mm}$		&	15.1		& 0.34 &\hfil $3.10$		&	65 \\ 
	DC28 (S2)&	\hfil $7.27$		&	14.9		& 0.31 &\hfil $6.29$		&	105 \\ 
	DC29	&	\hfil $7.04$		&	14.7		& 0.31 &\hfil $4.45$		&	70 \\ 
	DC30	&	\hfil $7.31$		&	14.7		& 0.31 &\hfil $4.26$		&	65 \\ 
	DC31	&	\hfil $4.51$		&	14.2		& 0.27 &\hfil $1.31$ 		&	110 \\ 
	DC32	&	\hfil $7.61$		&	14.2		& 0.32 &\hfil $5.92$		&	70 \\ 
	DC33	&	\hfil $8.78$		&	14.0  	& 0.34 &\hfil $4.00$		&	65 \\ 
	DC34	&	\hfil $10.9\hspace{2.8mm}$		&	14.0		& 0.37 &\hfil $4.26$		&	100 \\ 
	DC35	&	\hfil $8.98$		&	14.0		& 0.35 &\hfil $4.07$ 		&	65 \\ 
	DC36	&	\hfil $9.63$		&	13.8		& 0.36 &\hfil $6.08$		&	80 \\ 
	DC37	&	\hfil $9.53$		&	13.5		& 0.37 &\hfil $8.62$		&	100 \\ 
	DC38	&	\hfil $9.50$		&	13.0		& 0.38 &\hfil $5.71$		&	130 \\ 
	DC39	&	\hfil $10.1\hspace{2.8mm}$		&	12.6		& 0.40 &\hfil $6.17$		&	70 \\ 
	DC40	&	\hfil $11.7\hspace{2.8mm}$		&	11.8		& 0.44 &\hfil $3.51$		&	65 \\ 
	DC41	&	\hfil $11.4\hspace{2.8mm}$		&	11.1  	& 0.47 &\hfil $6.19$		&	65 \\ \hline
	\end{tabular}
	}
	\\ \\ Note. Column 1: ID of the DC halo.
	Column 2: mass of the DC candidate halo. 
	Column 3: the redshift. 
	Column 4: the virial radius.
	Column 5: the distance from the source halo. 
	Column 6: the time of the infall to the center of the source halo.
	Column 2-6 are the value when halos meet the DC criteria.
	\label{dc_table}	
\end{table}

\section{Results from the $N$-body Simulation}

\subsection{Star Formation and the LW Radiation} \label{sec_SF_LW}

In the zoom-in regions, the first Pop III star is formed at $z \simeq 35$.
The background LW radiation intensity
$J_{21}$ originating from Pop III stars 
($J_{21}^{\greeceIII}$) is peaked at $z \simeq 30$ but
gradually decreases afterward. 
Pop II sources first appear at $z \simeq 28$ and $J_{21}$ originating from
Pop II sources ($J_{21}^{\greeceII}$) grows rapidly to
the global mean of $\sim 10$ at $z\simeq 10$.

Figure~\ref{J21_evo} shows the distribution of $J_{21}$ evaluated  
at the centers of all the pristine halos at (a) $z = 23.1$ and (b) $9.3$. 
Except for the shift of the the peak intensity, 
the overall distributions
appear similar between the two epochs.
The distribution scales as $\propto J_{21}^{-1.6}$ for the
higher end, and the lower end decreases exponentially.
Our result is broadly consistent with that in \cite{Agarwal+2012}.
Interestingly, the slope at the higher end in our study is smaller than 
that of the Monte Carlo studies of 
\cite{Dijkstra+2008, Dijkstra+2014, Inayoshi+2015b},
who employ a two-point correlation function to populate galaxies.
The difference is likely caused by the assumed correlation at 
the small separation of $r \lesssim 1$~kpc. 
We have checked that the above features remain unchanged also for
more massive halos with masses greater than
$10^6$ and $10^7 \;M_\odot$.

We find that only the Pop II LW field
exceeds the critical values for DC.
Although a Pop III star is brighter in UV than a
Pop II star with the same mass, the global Pop III star formation 
rate density (SFRD) is overall small. 
For Pop III sources, moreover, the critical intensity
$J_{21}^{\rm crit}$ is higher than for Pop II sources (vertical
dashed lines in Fig.~1)
because of their higher effective temperatures
(equation~\ref{eq:jcr_pop}).
For these reasons, it is unlikely that the Pop III LW radiation
only causes DC events \citep[]{Agarwal+2012}. 
We thus consider mainly the LW radiation coming from Pop II stars.
We note here that $J_{21}$ varies in time and is
larger by an order of 
magnitude than $J^{\greeceII}_{21}$ when a DC cloud collapses.
This is because the luminous source (galaxy) attracts the DC cloud gravitationally,
as we discuss below in Section~\ref{sec_1100002}.

\subsection{DC Candidate Halos}
\begin{figure}[tb]
	\includegraphics[width=8.9cm]{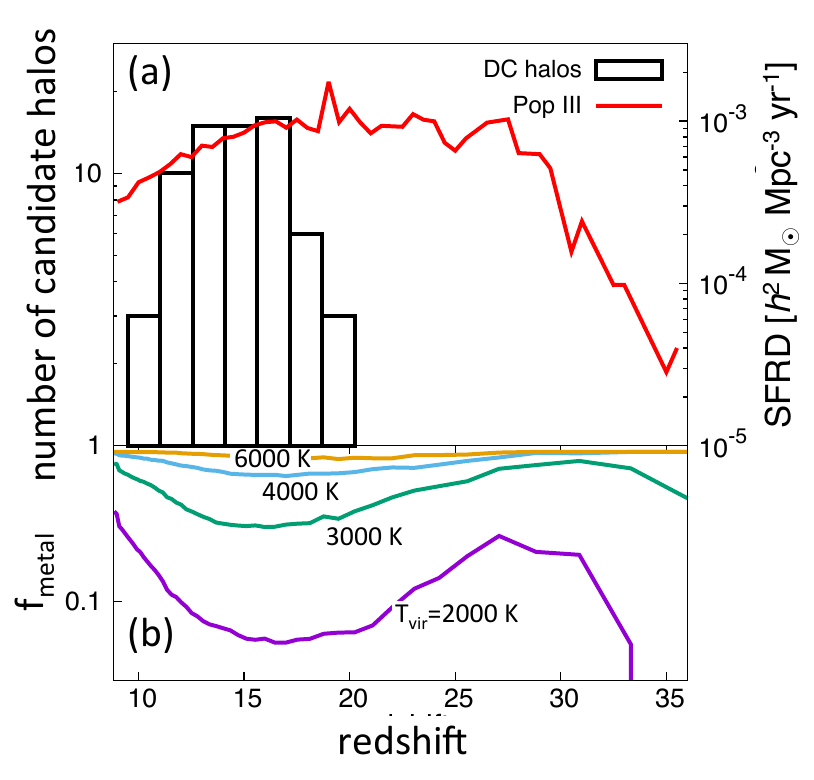}
	\caption{(a) Number of DC candidate halos (black) and the Pop III SFRD (red)
          as functions of redshifts for all the zoomed-in regions. 
	(b) Time evolution of the fraction of metal-enriched halos. Each line corresponds to the fraction for halos with 
	$T_\text{vir} = 2000$ (purple), $3000$ (green), $4000$ (blue), and $6000\;\mathrm{K}$ (yellow).}
	\label{dc_z_hist}
\end{figure}

\begin{figure}[tb]
	\centering
		\includegraphics[width=7.8cm]{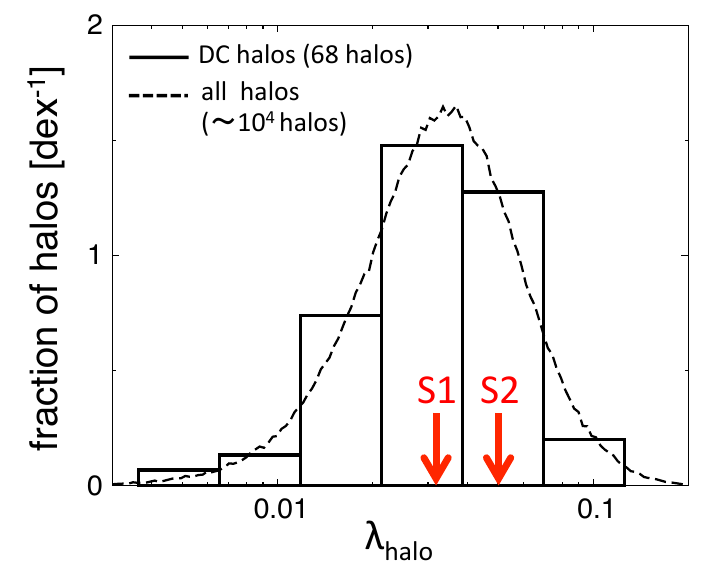}
		\caption{Spin parameter distributions for all the halos (dashed) and DC candidate halos (solid). 
		 The vertical axis represents the fraction of halos in each bin ($\mathrm{dex^{-1}}$).
		 Red arrows show the angular momentum for S1 and S2 halos,
		 which will collapse into the dense core (see Section \ref{HS}).
		 }
		\label{spin_param}
\end{figure}

In the 10 zoom-in regions, we find 68 DC candidates that satisfy all the
three conditions for DC given in Section~\ref{ssec:criteria}. 
Table \ref{dc_table} summarizes the properties of the
42 halos when each of them meets the DC criteria, whose hydrodynamical evolutions are further followed (Section 4).
Detailed evolutions of the halos labeled by F1, F2, S1, and S2 are described in 
Sections \ref{sec_1701848}, \ref{sec_1200071}, \ref{sec_1100002}, and \ref{sec_S2}, respectively.
Our hydrodynamics simulations show that only two cases
successfully collapse to trigger DC out of the 42 samples.
The two cases are dubbed as S1 and S2.

Figure~\ref{dc_z_hist}(a) shows the redshift distribution of the DC candidate
halos and the Pop III SFRD. 
The number of candidate halos is peaked at $z \simeq 15$ but distributed
widely over
$\Delta z \sim 10$, while the Pop III SFRD is almost constant from $z=30$ to $15$. 
The DC candidate halos appear at lower redshifts than Pop III stars
because strong LW radiation sources are needed for DC.
After $z=15$, the formation rate of DC candidate halos and the Pop III SFRD
decrease simultaneously, reflecting the fact that the number of newly
formed minihalos, the Pop III hosts, decreases
after $z=20$.

A similar trend can be also seen in Figure~\ref{dc_z_hist}(b), which shows
the fraction of metal-enriched halos with $T_\text{vir}=2000$, $3000$,
$4000$, and $8000\;\mathrm{K}$ (purple, green, blue, and yellow,
respectively).
Around $z = 15$, the fraction of metal-enriched halos starts
to increase for all the mass ranges and
even for the smallest halos that have $T_\text{vir} \simeq 2000\;\mathrm{K}$. 
Clearly, after $z = 15$, halos have fewer opportunities to grow up
until they attain $T_\text{vir} = 8000\;\mathrm{K}$ without being enriched with metals.
Metal enrichment causes the numbers of the DC candidate halos
and Pop III stars to decrease simultaneously at $z < 15$.

\subsection{Spins of DC Candidate Halos} \label{spin}
We examine whether the cloud collapse is affected by the spin
of its host halo. The degree of halo spin
is characterized by the nondimensional spin parameter defined by \cite{Bullock+2001}:
\begin{equation}
\lambda = \frac{J}{\sqrt{2}\;MV_\text{c}R_\text{vir}},
\end{equation}
where $J$ is the total angular momentum of the halo.
The distributions of the spin parameter for DC candidate halos and of all
halos are shown in Figure~\ref{spin_param}. 
The spin distribution can be well fitted by a lognormal distribution:
\begin{equation}
p(\lambda) = \frac{1}{\sqrt{2\pi} \sigma_\lambda} \exp{\left [ -\frac{\log ^2(\lambda / \bar{\lambda})}
{2\sigma_\lambda} \right ]} \frac{\mathrm{d}\lambda}{\lambda}.
\end{equation} 
The best-fit parameters for all halos are $(\bar{\lambda}, \sigma_\lambda) =  (0.034, 0.56)$, which is consistent with \citet{Bullock+2001} although the halo mass and redshift ranges
considered here are very different.

The spin parameter distribution for candidate halos also follows the lognormal distribution with
$(\bar{\lambda}, \sigma_\lambda) =  (0.043, 0.78)$ and the mean value and the variance of
the spin are consistent with those of all the halos. 

The red arrows in Figure~\ref{spin_param} show the spin parameters of the collapsed DC halos,
which host the cloud collapsing into the dense core (see Section \ref{HS}).
The marked values are comparable to the mean value of 
our detected DC candidate halos and also that of the all the halos in the simulation.
This indicates that angular momentum is not a critical factor
in the collapse process of DC halos.

\footnotetext{The absolute value of the Pop III SFRD of our simulation
  is higher than those found in the previous studies \citep{Agarwal+2012}. 
  This is likely owing to the fact that we only focus 
  on the zoomed-in regions, where star formation
  proceeds earlier.}

\section{Cloud Collapse in DC candidate Halos}
\label{HS}

We carry out hydrodynamical simulations to follow the subsequent
evolution of the gas clouds hosted by selected DC candidate halos. 
The calculations are performed for 42 halos out of the 68 samples extracted from
our parent $N$-body simulation. 
We find two ``successful'' cases, where the gas density at the cloud
center reaches 
$10^8 \;\mathrm{cm}^{-3}$ after evolving on an isothermal
track with $T\simeq 8000$ K. It is highly expected that a massive star, even
a supermassive star, is formed in such a gas cloud. 
For the other cases, however, the clouds do not collapse,
often because the host halo merges with a nearby light source halo
before the cloud collapses (see Sections \ref{sec_1701848} and
\ref{sec_1200071}).  
In this section, we first focus on the evolution of four
DC candidate halos, which are listed as F1, F2, S1, and S2 (DC14, DC2,
DC0, and DC28 in Table \ref{dc_table}, respectively). 
F1 and F2 are the failed cases, whereas S1 and S2 are successful ones. 
The overall features of the cloud evolution are 
summarized in Section \ref{sec_DC_hydro_g}.

\begin{figure}
	\centering
		\includegraphics[width=7.9cm]{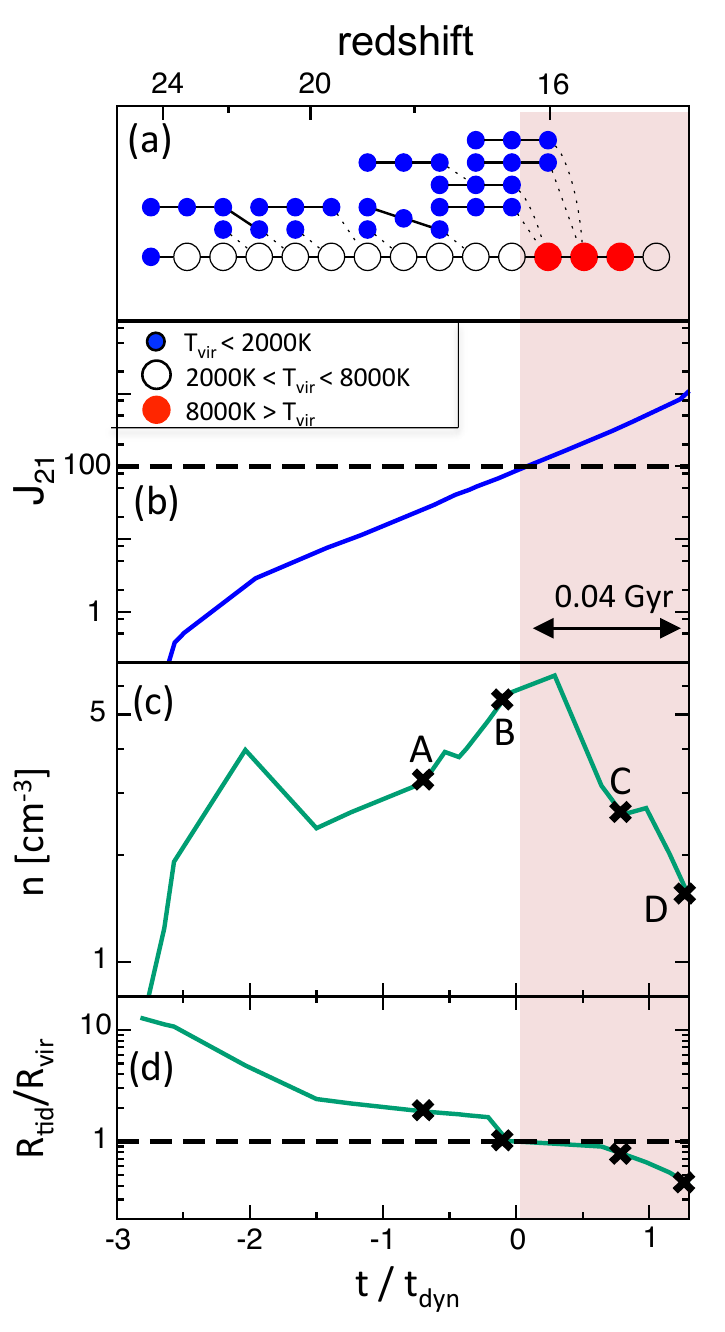}
		\caption{
(a) Halo merger tree for case F1 (DC14).
The small blue filled, black open, and red filled circles
represent halos with $T_\text{vir} < 2000\;\mathrm{K}$,
 $2000\; \mathrm{K} < T_\text{vir} < 8000 \;\mathrm{K}$, 
and $T_\text{vir} > 8000\;\mathrm{K}$, respectively. 
(b) Evolution of $J_{21}$ at the halo center. 
The solid line shows the evolution of the LW intensity,
and the dashed line indicates the critical LW intensity
necessary for the DC criteria, $J_{21}^\text{crit} = 100$.
(c) Time evolution of the maximum density within the cloud.
(d) Tidal radius normalized by the virial radius of the halo. 
		The dashed line indicates where the tidal radius becomes comparable to the virial radius of the host halo.
		In the shaded region, the virial temperature exceeds $8000$~K,
		with which the atomic H cooling becomes efficient.
		Points A, B, C, and D are the reference points used in Figure~\ref{halo_1701848}.
		The redshifts corresponding to $t/t_\text{dyn}$ are shown at the top of panel (a).
                The reference dynamical time at $t=0$ is $37.7\;\mathrm{Myr}$.
   		In panel (b), the black arrow represents the physical timescale of $0.04\;\mathrm{Gyr}$. 
}
		\label{mtree_1701848}
\end{figure}

\begin{figure*}[p]
	\centering
		\includegraphics[width=15cm]{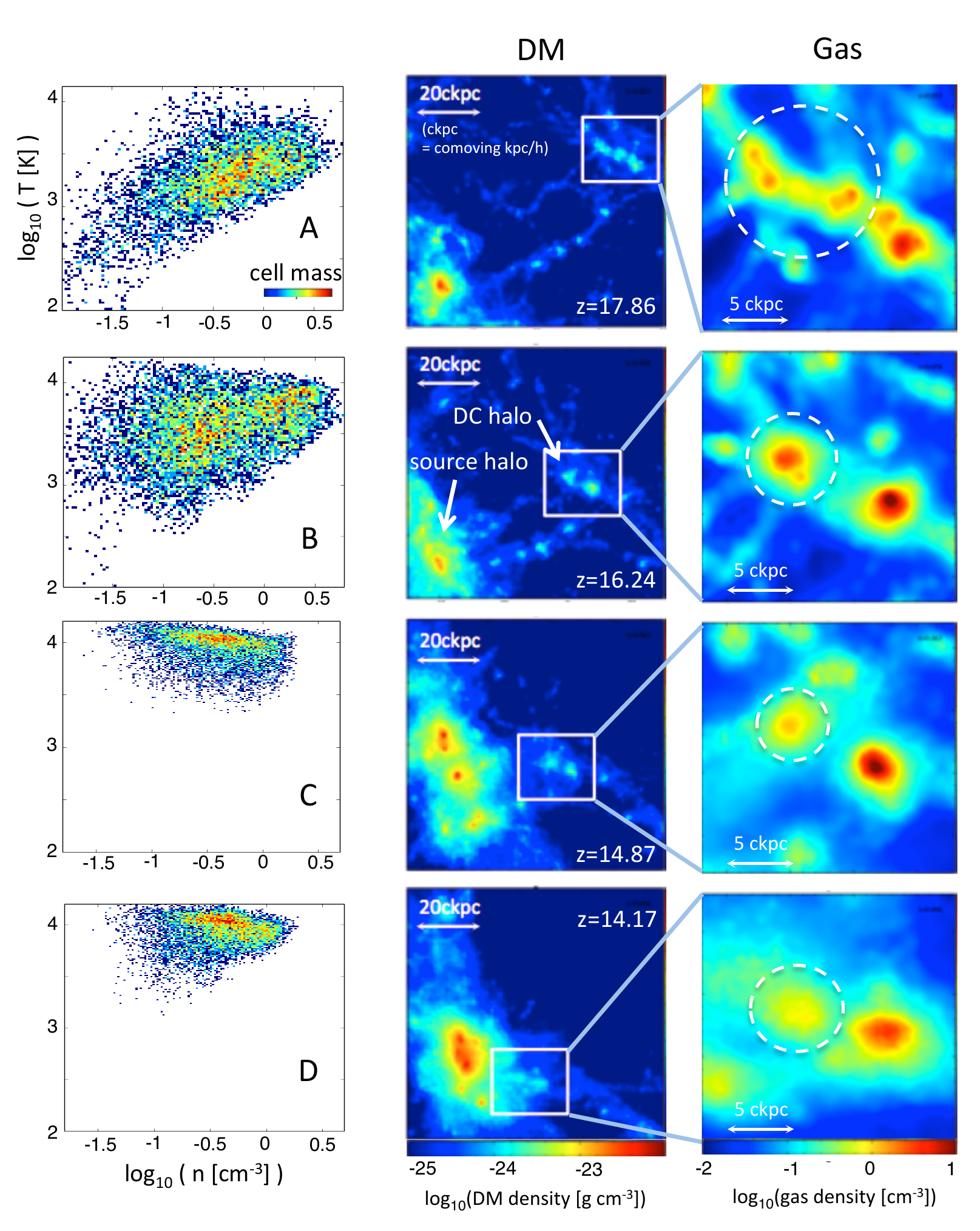}
		\caption{
		Time evolution of the F1 halo at four reference points shown in Figure~\ref{mtree_1701848}.
		Left:     the scatter plots of the temperature vs. the gas density. The color scale shows the cell mass in each bin.
		Middle: the projected DM density distribution around the F1 halo. 
		The large clump at the bottom left corner in each panel is the source halo. 
		The candidate halo is inside the white box. The color scale represents the DM density. 
		Right:   the projected gas density distributions around the halo.
		The color scale represents the gas density.
		There are two main clumps in each panel, and the clump on the left is the DC candidate halo, which is contained by 
		the white dashed circle. 
		The other clump is a halo in which a Pop III star is already formed. 
		In each panel, the comoving length scales of $20 \; \mathrm{kpc}/h$ and $5 \; \mathrm{kpc}/h$ 
		are represented with the white arrows. 
		}
		\label{halo_1701848}
\end{figure*}

\subsection{Case F1: Collapse Prevented by Strong Tidal Force} 
\label{sec_1701848}

We first study in detail the evolution of a DC candidate halo in which
DC does not occur (F1 in Table 1). 
Figures \ref{mtree_1701848}(a) and (b) show the merger history of F1
and the evolution of the LW intensity at the halo center, respectively. 
LW radiation turns on around $z \simeq 24$ when 
the halo virial temperature just reaches $T_\text{vir} = 2000\;\mathrm{K}$.
The intensity increases gradually and reaches the critical value
$J_{21}^\text{crit}=100$ at  $z \simeq 16.5$. 
Pop III star formation in the halo is suppressed by the strong 
LW radiation at $16.5 < z < 24$, when the halo virial temperature is in the range 
$2000\;\mathrm{K} < T_\text{vir} < 8000\;\mathrm{K}$. 
Note that $\text{H}_2$ cooling would operate otherwise
in this temperature range.
The halo acquires its mass through smooth accretion and minor mergers,
but no major mergers are observed.\footnote{We define a major merger
as one with a mass ratio $m_1/m_2 < 3$, where $m_1$ is the mass of the larger halo.
Otherwise, the merger is defined as a minor merger.} 

The gas density reaches $5\;\mathrm{cm^{-3}}$ at $t = 0$
when the halo satisfies the DC criteria (Fig.~\ref{mtree_1701848}c).
However, the gas does not continue contracting, and the density starts to decrease afterward.
The halo merges with its nearby light source halo at $t \sim 1.3t_\text{dyn}$,
before the gas cloud goes gravitational collapse.

The nearby light source halo affects not only 
the abundance of $\text{H}_2$ in the DC halo by photodissociation,
but also the dynamical evolution through tidal forces.
The tidal radius of a halo, $R_\text{tid}$, outside which the matter will be
gravitationally disrupted, is defined by
\begin{equation} \label{eq_tidal}
R_\text{tid} = \left ( \frac{M_\text{halo}}{3M_\text{source}} \right )^{1/3} d_\text{dist},
\end{equation}
where $M_\text{halo}$ is the mass of the candidate halo, $M_\text{source}$ is the source
halo mass that exerts the external gravitational force, and $d_\text{dist}$ is
the distance between the two halos \citep{Binney&Tremaine1987}. In reality, the above
expression is modified by a minor factor when the rotation of the candidate halo around
the source halo and the density profile of the source halo are taken into account.
These complications, however, introduce only a small change less than a factor of 2
for $R_\text{tid}$ \citep{Binney&Tremaine1987}. We thus use eq. \eqref{eq_tidal}
in the following discussion.

Figure~\ref{mtree_1701848}(d) shows the time evolution of $R_\text{tid}$ normalized
by the virial radius of the candidate halo, $R_\text{vir}$.  The dashed horizontal
line shows where $R_\text{tid}$ is equal to $R_\text{vir}$, that is, the matter
at $R_\text{vir}$ in the DC halo can be stripped by the strong tidal field.
After epoch B, when $R_\text{tid} \sim R_\text{vir}$, the density begins to decrease
because the gas cloud is disrupted rather than collapses.

In Figure~\ref{halo_1701848}, the panels in each row show the 
snapshots at the epochs A--D indicated
in Figure~\ref{mtree_1701848}(a). 
The panels in the left column show that the temperature of the central
part of the cloud reaches $10^4 \;\mathrm{K}$ at epoch 
B. The snapshots C and D show that most of the gas has a temperature 
around $10^4\;\mathrm{K}$. 
The panel sequence shows that the candidate halo indicated by the square
is attracted by the light source halo through epochs A--D.
At D, the candidate halo is completely merged 
with the source halo, together with its surrounding matter.
The right panel (D) also shows that the candidate halo and the gas
cloud itself become elliptical in shape and
elongated toward the source halo. 
This is a characteristic feature of tidal disruption, stretching
an object toward the external gravitational source. 
Clearly, the density decrease is caused by the tidal field 
of the massive light source halo.

The candidate halo experiences multiple mergers with less massive halos
after the virial temperature reaches $8000\;\mathrm{K}$ (Fig.~\ref{mtree_1701848}a). 
The infalling halos carry energy into the center of the candidate halo
if they survive to reach the center, in a similar manner to
dynamical disk heating \citep{Toth&Ostriker1992}. Once the energy is deposited
to the center of the candidate halo, the center starts to expand
and the density decreases. Then the cloud becomes more susceptible
to tidal disruption. The expanded core is only loosely
bounded gravitationally, and the gas density decreases further
(Fig.~\ref{halo_1701848} B).
Interestingly, this is a two-step process;
minor mergers under the strong tidal field initiate
the destruction of the cloud. 
The density decrease at $z \simeq 22$ ($t \sim -2 t_\text{dyn}$) is
caused by a minor merger.

As we will discuss in Sections \ref{sec_1100002} and \ref{sec_DC_collapse},
major mergers can help induce gravitational
instability at the cloud center. 
The difference from the above discussion is simply whether or not the merger assembles
and brings enough mass toward the cloud center.

\begin{figure}[tb]
	\centering
		\includegraphics[width=8cm]{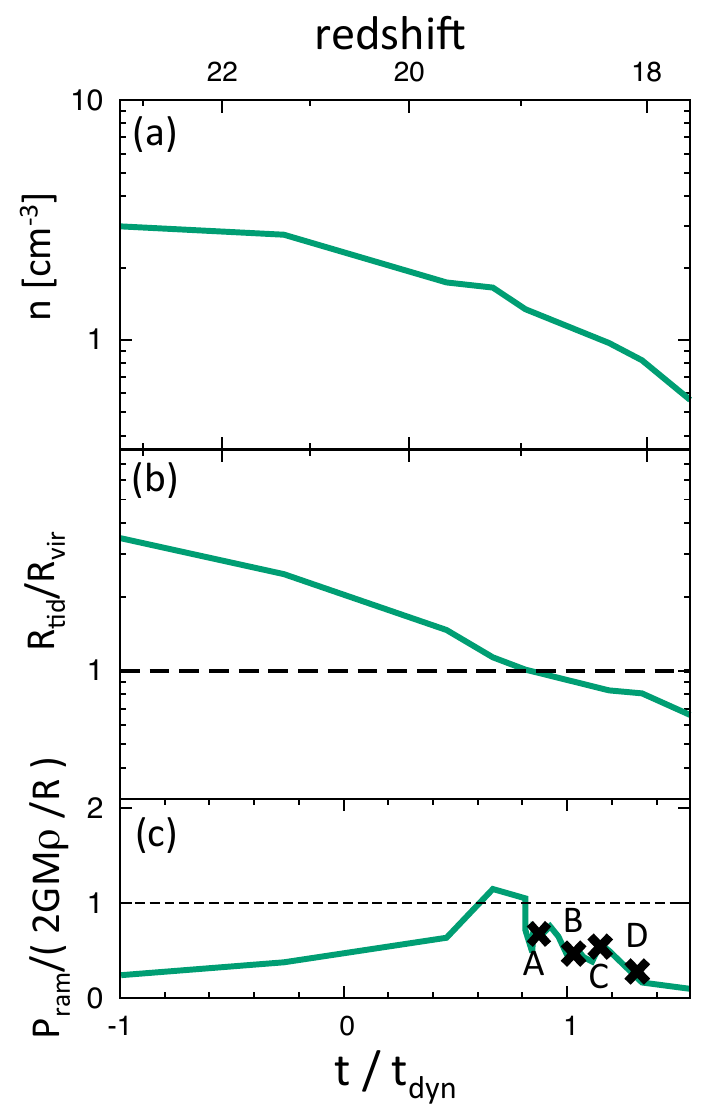}
		\caption{
(a) Evolution of the gas density in the DC cloud 
for case F2 (DC2). 
The horizontal axis represents the elapsed time since
the host halo satisfies DC criteria 
and normalized by the dynamical time at $t = 0$.
(b) Evolution of the tidal radius normalized 
by the virial radius of the host halo.
The dashed line indicates where the tidal radius 
becomes equal to the virial radius of the host halo.
(c) Ratio of the ram pressure exerted on the cloud (eq. \ref{eq_Pram})
and the gravitational force per unit area. 
The dashed line represents where the ram pressure and the gravitational force balance. 
The points A, B, C, and D are the reference points used in Figure~\ref{halo_1200071}.
The reference dynamical time at $t=0$ is $t_{\rm dyn} = 27.7\;\mathrm{Myr}$.
		}
		\label{profile_1200071}
\end{figure}

\begin{figure}[tb]
	\centering
	\includegraphics[width=9cm]{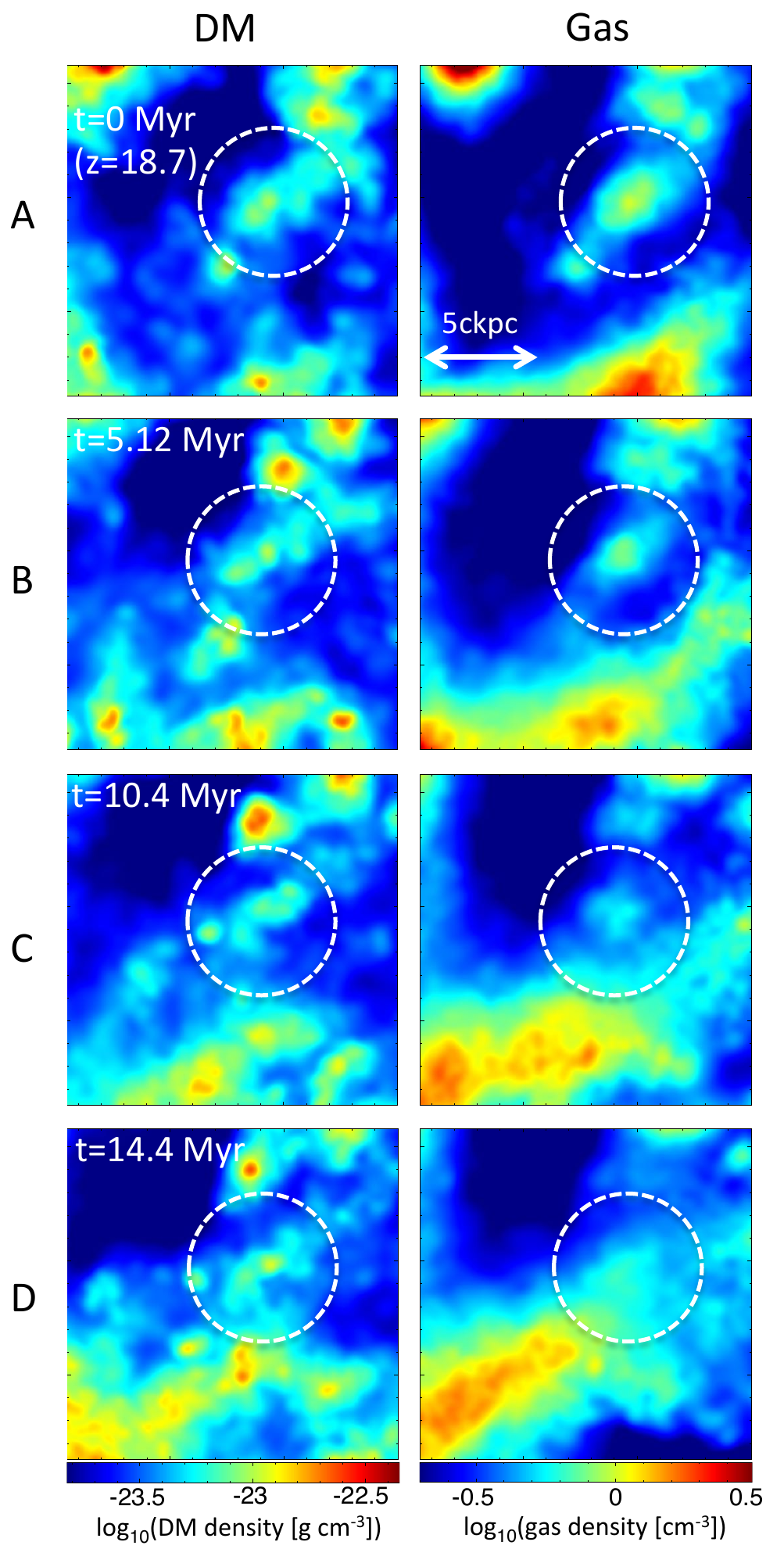}
	\caption{
	Time evolution of F2 halo at the four reference points shown in Figure~\ref{profile_1200071}(c).
	Left: the projected DM density distributions. 
	The large clump at the top left corner in each panel is the source halo. 
	The candidate halo is inside the white dashed circle. 
        The color scale represents the DM density. 
	Right:   the projected gas density around the candidate halo, the same region as in the left column. 
	The color scale represents the gas density.
	The elapsed time $t$ since the epoch of panels A is attached to each panel.
	 }
	\label{halo_1200071}
\end{figure}

\begin{figure}[tb]
	\centering
	\includegraphics[width=7.85cm]{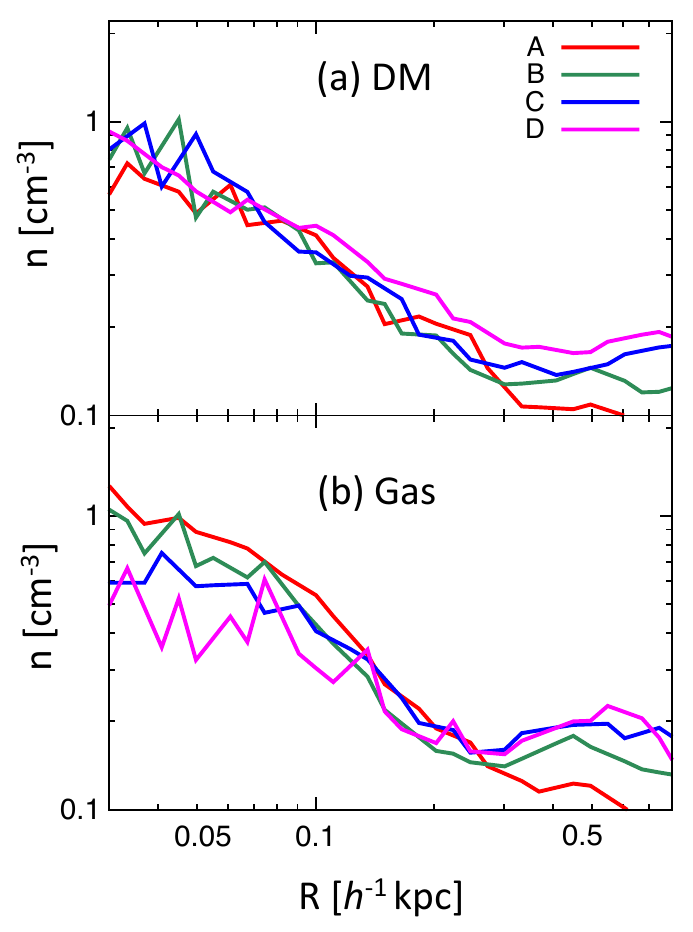}
	\caption{
	Density profiles of the (a) DM and (b) baryons for the F2 halo at the four reference points shown in Figure~\ref{profile_1200071}(c). 
	The horizontal axis shows the distance from the density maxima of the cloud. 
	Here the DM density is converted to the corresponding baryon density by 
	$n_\text{DM} \equiv f_\text{b}\rho_\text{DM}/(1.2 m_\text{p})$, where $1.2$ is the mean molecular weight of the neutral gas. 
	 }
	\label{rho_profile_1200071}
\end{figure}

\subsection{Case F2: Collapse Prevented by Ram Pressure Stripping} 
\label{sec_1200071}

We also find that ram pressure often prevents cloud collapse.
Figures \ref{profile_1200071}(a) and (b) show the evolution 
of the density and the tidal radius, respectively. 
Even after the virial temperature of the halo exceeds $8000\;\mathrm{K}$ 
at $t = 0$, the density decreases continuously. 
The tidal force of the light source halo
becomes comparable to the gravitational force at $t \sim t_\text{dyn}$. 
The halo experiences neither minor nor major mergers for 
$\simeq 50 \;\mathrm{Myr}$ after the halo satisfies the DC criteria.

We suspect that hydrodynamic effects cause the gas cloud to
disperse. Ram pressure is given by the product of density $\rho_\text{gas}$
and velocity $v$ as
\begin{equation}
P_\text{ram} \equiv \rho_\text{gas} v^2. \label{eq_Pram}
\end{equation}
If a cloud is moving in a dense medium, ram pressure acts
on the cloud to strip the outer gas against its self-gravity.
The condition for the stripping is given by
\begin{equation}
P_\text{ram} > \alpha \frac{G M(R) \rho_\text{gas}}{R}
\end{equation} 
\citep{Gunn&Gott1972,Gisler1976},
where $R$ is the size of the cloud, $M(R)$ is the gravitating mass within $R$
including DM, and $\alpha$ is an order unity parameter that is characterized
by the geometry of the cloud. According to the detailed simulations
by \cite{McCarthy+2008}, $\alpha = 2$ is appropriate for a spherical cloud.
We find that the ram pressure acting on the cloud is comparable to the gravitational force 
at $t \sim 0.6t_\text{dyn}$ (Fig.~\ref{profile_1200071}c).  

The discrepancy between the DM and gas profiles exhibits another evidence
for the ram pressure stripping acting on the cloud.
Figure~\ref{halo_1200071} shows the projected density distributions for
DM and baryons at the reference points A, B, C, and D in
Figure~\ref{profile_1200071}(c).  
The cloud is moving toward the dense filament at the
bottom. On its way, the cloud is dissolved into the surroundings, 
while the DM structure remains intact. Clearly, the candidate halo does
not experience mergers.
Figure~\ref{rho_profile_1200071} shows the density profiles of 
the (a) DM and (b) gas. The gas density 
decreases with time, while the DM profile remains almost unaffected. 

\begin{figure}[tb]
	\centering
	\includegraphics[width=8.2cm]{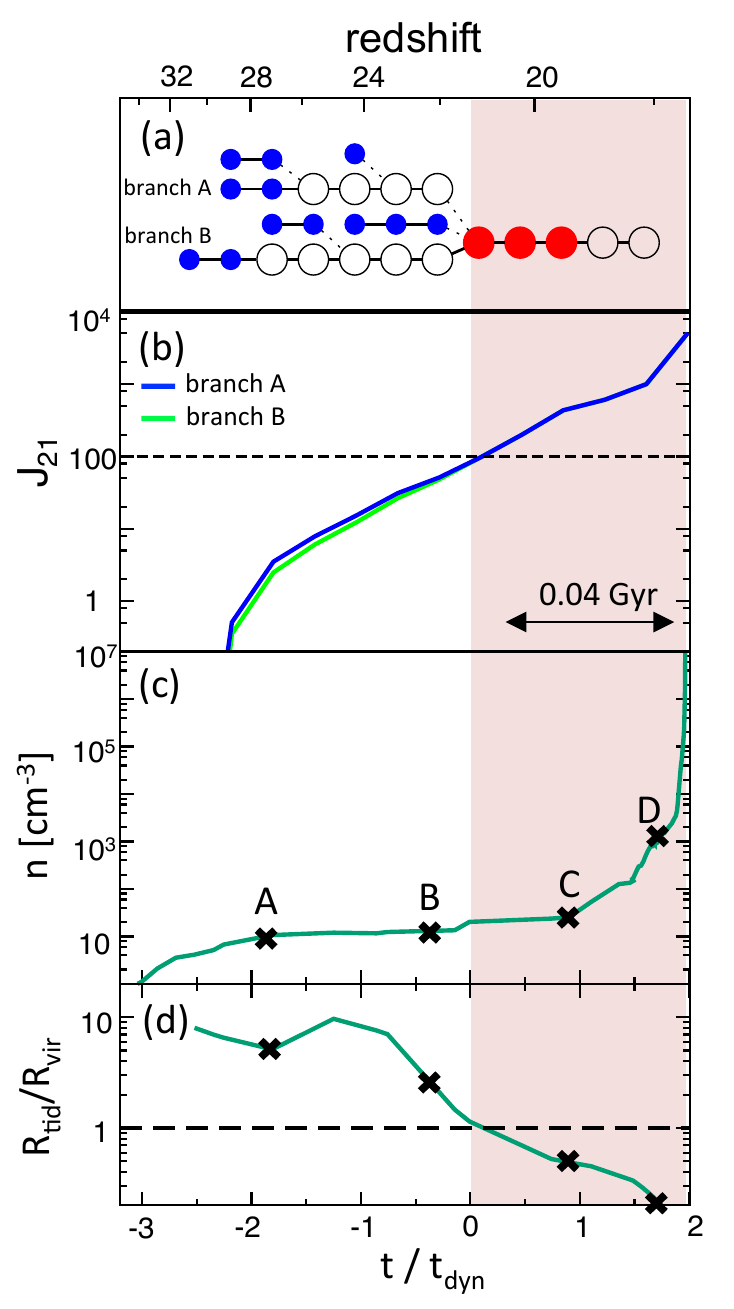}
	\caption{
	Same as Figure~\ref{mtree_1701848} but for case S1.
	There are two main branches (A and B) in the merger tree, and 
	the DC candidate halo is the product of the major merger of these two branches.
	The solid blue and green lines in panel (b) show the evolution of the LW intensity at the halo center of branches A and B.
        The reference dynamical time at $t=0$ is $t_{\rm dyn} = 26.9\;\mathrm{Myr}$.
	}
	\label{mtree_1100002}
\end{figure}

\begin{figure}[tb]
	\centering
		\includegraphics[width=7.85cm]{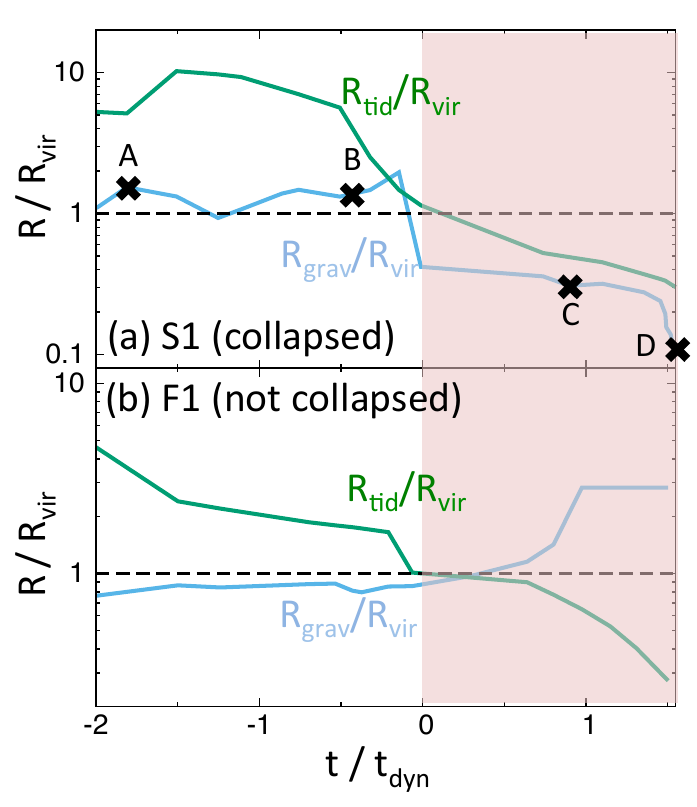}
		\caption{
		Evolution of gravitational radius (eq. \ref{m_BE}) relative to the virial radius (blue) and
		the tidal radius relative to the virial radius (green) for (a) S1 and (b) F1.
		The time origin $t=0$ is set to be the moment at which the halo satisfies the DC criteria, 
		and the time is normalized by the dynamical time at $t=0$.
		}
		\label{rgrav_1100002}
\end{figure}

\begin{figure*}[p]
	\centering
		\includegraphics[width=15cm]{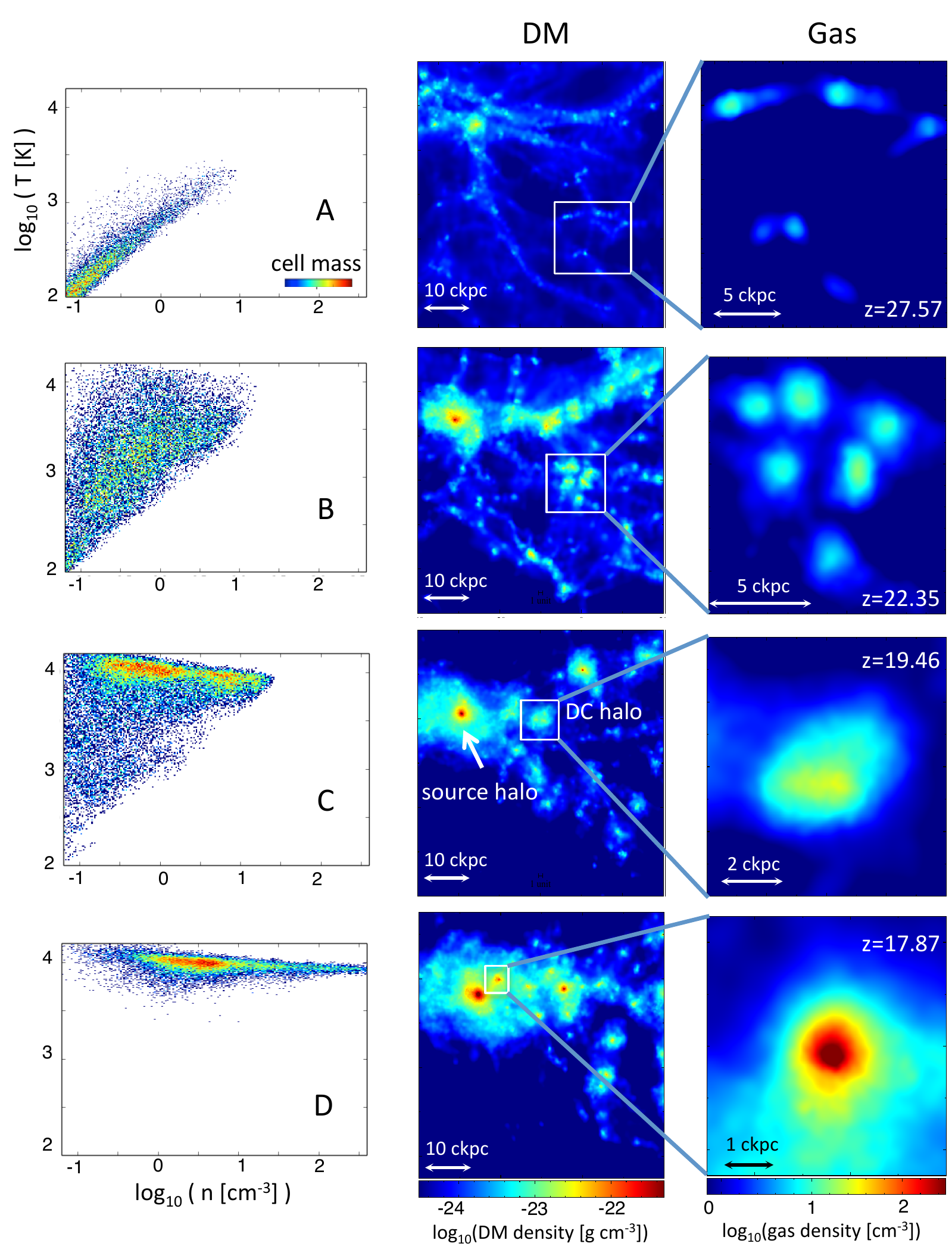}
		\caption{
		Same as Figure~\ref{halo_1701848}, but for four reference points for case S1 shown in Figure~\ref{mtree_1100002}.
		There are six small clumps in epochs A and B tha are merged into the DC candidate halo in epochs C and D.
		}
		\label{halo_1100002}
\end{figure*}

\begin{figure}[tb]
	\centering
		\includegraphics[width=8.cm]{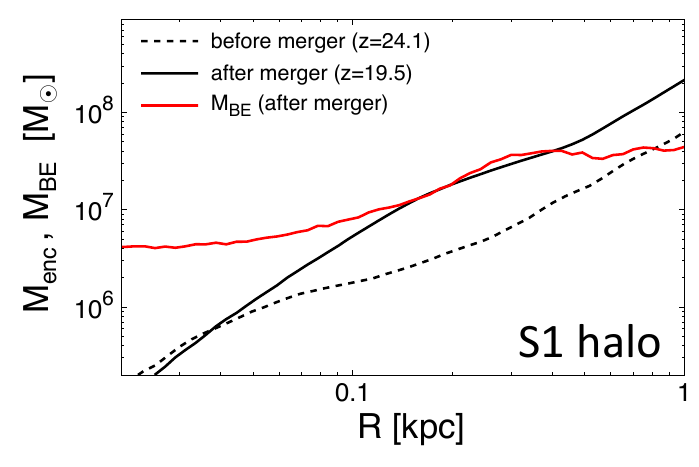}
		\caption{Radial profiles of the enclosed mass before ($z=24.1$; black dashed) and after the major merger ($z=19.5$; black solid).
		The red solid line shows the profile of the Bonnor--Ebert mass after the merger.
		}
		\label{Menc_1100002}
\end{figure}

\subsection{Cases S1 and S2: DC Clouds}
\label{sec_1100002}

In cases S1 and S2, the gas cloud continues to collapse 
and an SMS is likely to form. 
We mainly discuss the evolution for case S1 below, 
while the S2 case is briefly discussed in Section \ref{sec_S2} 
because its overall evolution is quite similar to that of S1.

\subsubsection{The evolution of S1}

The merger history of halo S1 and the evolution of the LW
intensity at its progenitor center are shown in
Figures~\ref{mtree_1100002}(a) and (b). 
There are two main tree branches (A and B), which are finally
merged into the candidate halo.
Progenitors in each branch are exposed to strong radiation with
nearly the same LW intensity,
which suppresses Pop III star formation even when $T_\text{vir} > 2000\;\mathrm{K}$.
The halo temperature increases from $T_\text{vir}=2000$ to $8000\;\mathrm{K}$
in $\simeq 50 \;\mathrm{Myr}$, which is shorter by a factor of 2
than the same time duration for halo F1 (Section \ref{sec_1701848}).
The rapid growth of S1 is enabled by the major merger of two halos A and B, at $t \sim 0$. 
Figure~\ref{mtree_1100002}(c) and (d) show that the density increases
monotonically and that the tidal radius becomes equal to the virial
radius of the halo right after the major merger. 
The cloud is able to collapse even under the strong tidal field. 

Gravitational collapse can be induced not only by the self-gravity
of the gas cloud but also by the gravity of the DM halo.
The DM density profile of the candidate halo,
$\rho_\text{DM}(r)$, follows a simple relation with the gas density $\rho(r)$
as $\rho_\text{DM}(r) = \Omega_\text{m}/ \Omega_\text{b} \rho(r)$ when the density is
less than $100\;\mathrm{cm}^{-3}$ \citep{Choi+2015}.
The radius  $R_\text{grav}$
of an isothermal cloud that experiences gravitational
collapse is estimated by the ratio of
the enclosed mass to the local Bonnor--Ebert mass, defined as follows:
\begin{equation} \label{m_BE}
M_\text{BE} \equiv \frac{1.18\; f_\text{b}^{1/2}c_\text{s}^4}{p^{1/2}G^{3/2}} \ge \int ^{R_\text{grav}} _0 \left [ \rho_\text{DM}(r) + \rho(r) \right ] 4\pi r^2\mathrm{d}r,
\end{equation}
where $c_\text{s}$ is the local sound speed, 
$f_\text{b} \equiv \Omega_\text{b}/ \Omega_\text{m}$,
and $p$ is the pressure of the cloud. The factor $f_\text{b}$ explains
the effect of the DM potential. In the derivation of eq. \eqref{m_BE},
the gravitational force originating from the matter outside the virial
radius of the halo is neglected.
We use the Bonnor-Ebert mass because the cloud is approximately static
before the onset of the collapse.

Figure~\ref{rgrav_1100002}(a) compares the evolution of gravitational
radius $R_\text{grav}$ and tidal radius $R_\text{tid}$. 
Just before $t \sim 0$, $R_\text{grav}/R_\text{vir}$ drops almost by
an order of magnitude.
When $R_\text{grav}$ is much smaller than $R_\text{tid}$, the cloud can
collapse without being disrupted by the light 
source halo. In the case of F1, $R_\text{grav}$ becomes larger 
than $R_\text{tid}$ at $t \sim 0$, so that the cloud cannot collapse 
because of the strong tidal force (Fig.~\ref{rgrav_1100002}b).

The right panels of Figure~\ref{halo_1100002} exhibit how successive
mergers take place to lead to DC. Six clumps are finally merged into a clump at
C. Before the merger, the mean temperature of the clouds is
$\simeq 4000 \; \mathrm{K}$ (left column of Fig.~\ref{halo_1100002}). 
It finally settles at $\simeq 8000\;\mathrm{K}$ at C. 
The rapid temperature increase is caused by the strong shock generated
by the merger and enables efficient atomic hydrogen cooling.
Radiative cooling then reduces the pressure support,
promoting the collapse of the cloud. Interestingly, the two clumps merge
at a relative velocity of $23.0 \;\mathrm{km\;s^{-1}}$,
which is nearly twice the sound speed of the cloud. 

The major merger also concentrates matters toward the halo center.
Figure~\ref{Menc_1100002} shows the total enclosed mass within
the distance $R$ from the halo center before (dashed) and after
the merger (black solid). The mass increases by an order of magnitude
at $R \sim 0.1 \;\mathrm{kpc}$. 
After the merger, the enclosed mass exceeds
$M_\text{BE}$ (red curve in Fig.~12)
at $R \sim 0.1 \;\mathrm{kpc}$.
This induces the gravitational instability. 
Thus, the major merger also causes
a sudden decrease of $R_\text{grav}$ by an order
of magnitude (Fig.~\ref{rgrav_1100002}).

Major mergers increase the central density of the cloud,
while the minor mergers act in the opposite way as
shown in Section~\ref{sec_1701848}. 
Within the cloud core, the gravitational force dominates over
the pressure force and over the tidal forces exerted by
neighboring halos.
Apparently the mass of the extended core determines whether the
cloud expands or contracts. We will discuss this issue more
in detail in Section \ref{sec_DC_collapse},
where we quantify the core entropy as a proxy to
the collapse criteria.

\subsubsection{The evolution of  S2} \label{sec_S2}
The cloud in S2 also collapses in essentially the same manner as S1.
There are two notable differences, however. One is that the major merger
occurs about one dynamical time after when the halo satisfies the DC criteria.
Hence the merger acts as promoting the collapse.
The other is that the closest light source halo has a similar mass to S2.
It exceeds the atomic-cooling threshold $20 - 30\;\mathrm{Myr}$ earlier than S2
itself. Apparently our S2 is a similar system to the ``synchronized pair'' studied in
\cite{Visbal+2014b}, 
where the source halo and the main DC halo become atomic-cooling halos simultaneously.
In the case of S2, there is another luminous halo, which contributes
more than half of the total $J_{21}$ at S2.

The mass of the closest light source halo is not large enough
to cause significant tidal effects on the main DC halo.
We find that $R_\text{vir}$ remains smaller than $R_\text{tid}$ 
over one dynamical time after the halo satisfies the DC criteria.
The central density remains high and the cloud 'survives'
against disruption until the final merger triggers DC.

\begin{figure*}[!htbp]
	\centering
	\includegraphics[width=16.2cm]{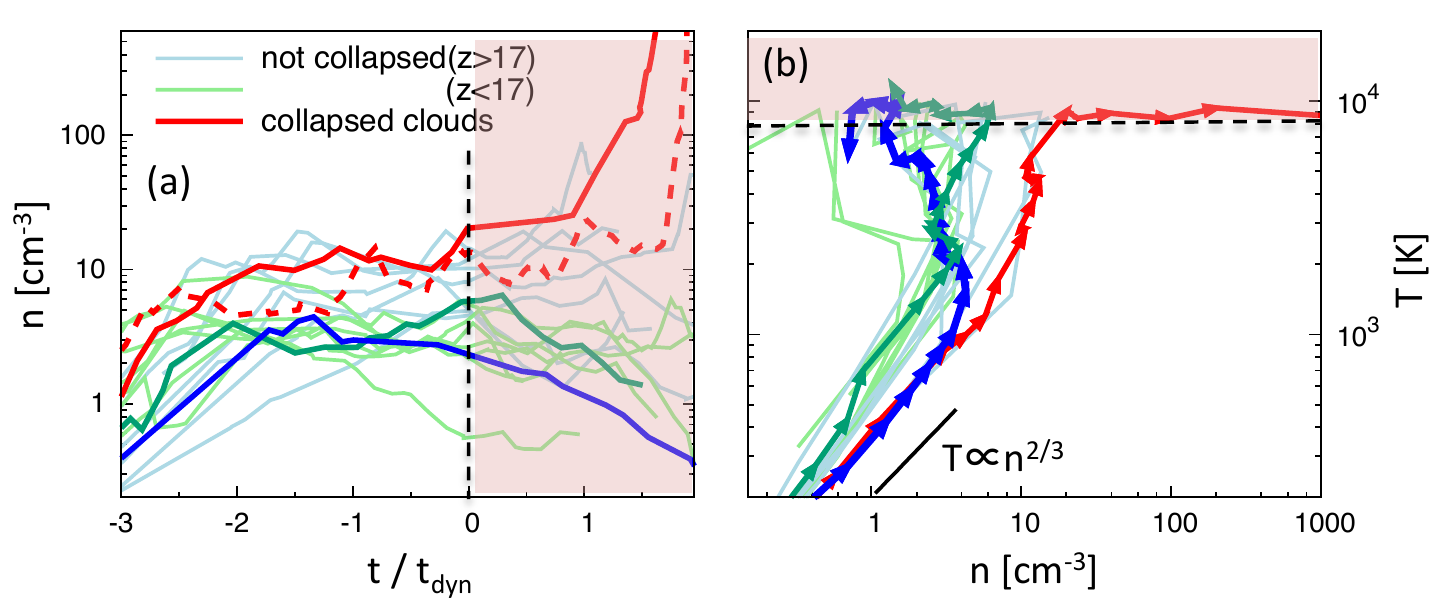}
	\caption{
Time evolution of 16 gas clouds hosted by DC candidate halos.
(a) Evolution of the central density $n$ as functions of time $t$.
We mark the four representative cases with thick lines: F1 (green), F2 (blue), 
S1 (red solid), and S2 (red dashed).	
The blue and green lines represent the uncollapsed clouds 
that satisfy the DC criteria at $z > 17$ and 
at $z < 17$, respectively. 
At the end of each line, the target halo is merged with the 
light source halo, except for the collapsed clouds.
(b) Temperature evolution as a function of the density 
for the 15 representative cases (S2 is not shown because it follows 
almost the same path as for S1). 
The vectors denote the evolution during every $0.25 t_\text{dyn}$
for cases F1, F2, and S1 as in panel (a).
For the other 12 DC clouds, we only show the evolution paths on the
plane, and the evolution is smoothed out for every $t_\text{dyn}$. 
The black solid line shows the adiabat, $T \propto n^{2/3}$.
In each panel, the shaded region indicates $T_{\rm vir} > 8000\;\mathrm{K}$, 
where atomic hydrogen cooling becomes efficient.
}
\label{rho_T_evo}
\end{figure*}

\begin{figure}[tb]
	\centering
		\includegraphics[width=8.5cm]{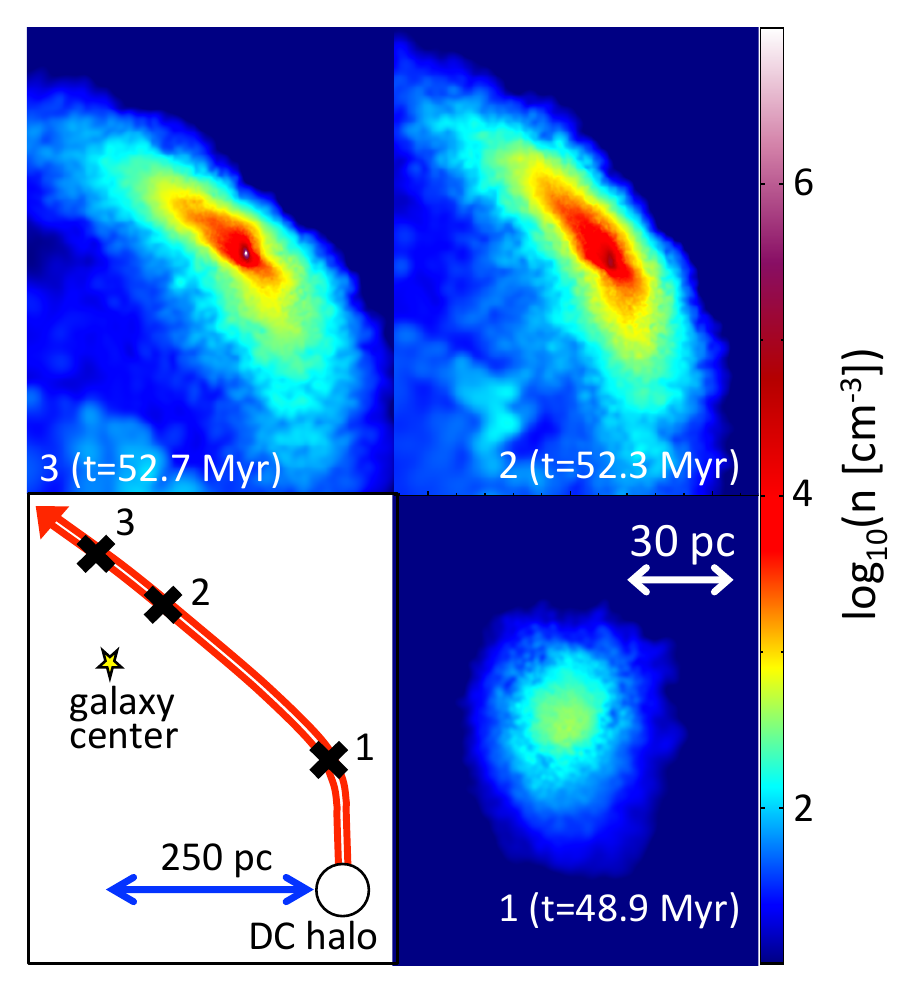}
\caption{
Bottom left: the schematic picture of the locus of the S1 halo relative 
to the source halo. The red arrow shows the path of the DC halo relative 
to the light source halo. Bottom right, top right, top left: The projected gas density map around 
the DC halo. The panels show the cloud's snapshots at $t=48.9$, $52.3$, 
and $52.7\;\mathrm{Myr}$ counterclockwise, corresponding to the points 
1, 2, and 3 in the bottom left panel. 
The time origin represents the epoch when
the halo meets the DC criteria.
}
\label{collapse_1100002}
\end{figure}

\begin{figure}[tb]
	\centering
		\includegraphics[width=8.05cm]{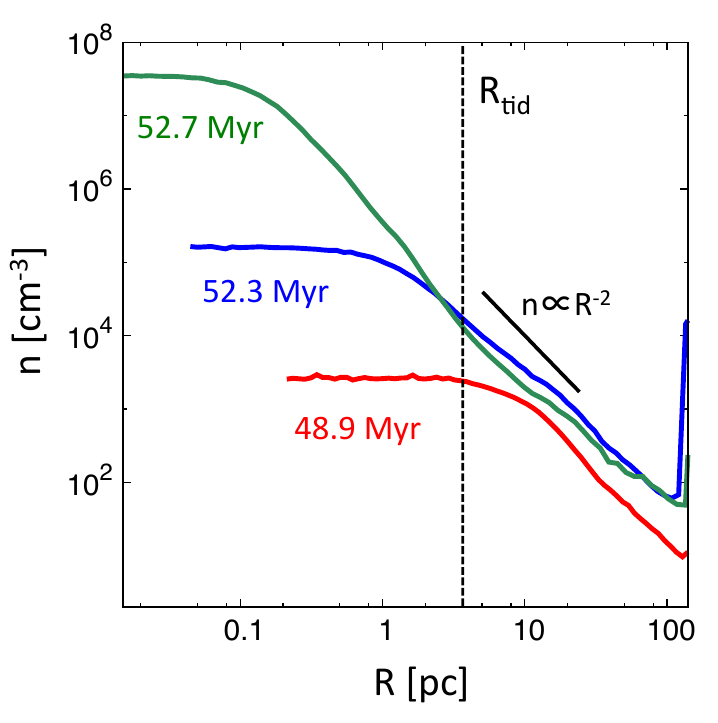}
\caption{Radial profiles of the gas density in the S1 halo 
at $t=48.9$, $52.3$, and $52.7\;\mathrm{Myr}$. 
The dashed line indicates the tidal radius at which the tidal 
force exceeds the self-gravity of the gas cloud at $t=52.7\;\mathrm{Myr}$. 
}
\label{rho_r_1100002}
\end{figure}

\begin{figure}[tb]
	\centering
		\includegraphics[width=7.8cm]{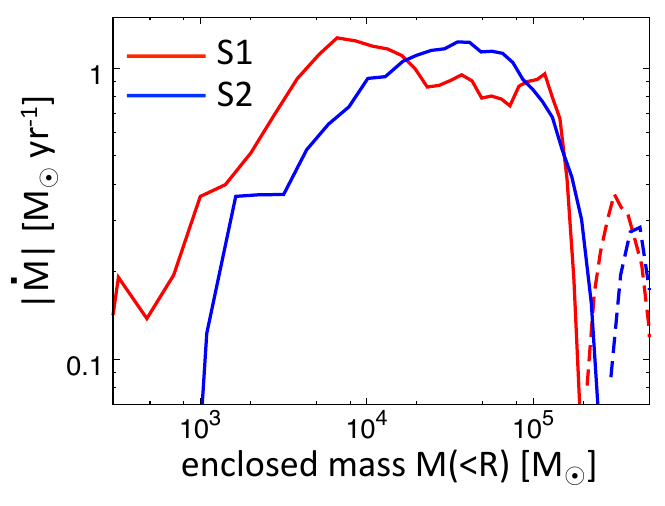}
		\caption{
		Expected mass infall rates $|\dot{M}| = 4\pi r^2\rho |v|$
		as functions of the enclosed mass $|M(<R)|$.
		The snapshots when the central density reaches $2\times10^7\;\mathrm{cm}^{-3}$
		are used to define the profiles. 
		We plot the profiles for S1 (red) and S2 (blue).
		The solid lines show the infall rates, while the dashed lines show the outflow rates.
		}
		\label{Mdot_1100002}
\end{figure}

\subsection{Thermal Evolution of DC Gas Clouds} 
\label{sec_DC_hydro_g}

We aim at deriving general features in the evolution of DC clouds.
Figure~\ref{rho_T_evo}(a) shows the density evolution of the DC candidate halos. 
We plot the evolution until each host halo merges with its nearby light source halo.
For the low-$z$ samples (green lines) that satisfy the DC criteria at $z<17$,
the density peaks at $n = 1-3 \;\mathrm{cm^{-3}}$ at $t \sim 0$, 
but gradually decreases afterward. 
The high-$z$ samples (blue lines) at $z > 17$ have relatively higher
densities than the low-$z$ samples owing to the cosmic expansion. 
By performing high-resolution hydrodynamic simulations,
we are able to robustly determine whether or not DC
is triggered in the candidate halos.

Figure~\ref{rho_T_evo}(b) shows the temperature evolution as a function
of density. At the beginning of the
collapse, the density and temperature evolve adiabatically as $T \propto n^{2/3}$.
The temperature rises monotonically to 
$T \simeq 8000 \;\mathrm{K}$, tracing roughly the halo virial temperature.
At $T > 1000\;\mathrm{K}$, the evolution deviates upward
from the adiabatic track because of dynamical heating associated
with mergers.
The entropy monotonically increases because radiative cooling
is inefficient until the temperature reaches $8000\;\mathrm{K}$.

The density of the uncollapsed cloud reaches only 
$\sim 10 \;\mathrm{cm}^{-3}$. 
The Jeans length at the maximum density and at $T = 8000\;\mathrm{K}$ is
$10^2 - 10^3\;\mathrm{pc}$ ,
which is comparable to the virial radius of the host halo. 
In the case of F1, the tidal force at the virial radius
becomes very strong because 
the DC halo approaches the light source halo rapidly before
the onset of the collapse. 
We will show that similar evolution is seen in many of
the other uncollapsed cases in Section \ref{sec_tinf_sim} below.

\section{Formation of a protostar} \label{proto_star_1100002}

For S1 and S2, we continue the calculations until
the central density reaches $10^{8} \;\mathrm{cm^{-3}}$.
In both the cases, the target gas cloud approaches quickly,
while contracting, to the light source halo.
The bottom left panel of Figure~\ref{collapse_1100002} shows a
schematic picture of the orbit of the S1 halo relative to
the light source halo. The cloud is attracted gravitationally
and passes by the light source.
Figure~\ref{collapse_1100002} also shows the projected gas density
around the DC halo at $t=48.9$, $52.3
$, and $52.7\;\mathrm{Myr}$.  
The bottom right panel ($t=48.9 \;\mathrm{Myr}$) corresponds to the
reference point D in Figure~\ref{mtree_1100002}. 

The DC conditions are satisfied at $t=0$. 
At $t=48.9\;\mathrm{Myr}$, the cloud is almost spherical while it is
tidally distorted at $t = 52.3\;\mathrm{Myr}$. 
The cloud approaches $100\;\mathrm{pc}$ from the light source halo at
$t=52.3\;\mathrm{Myr}$.
During the passage of the pericenter, the density
increases and reaches $10^8\;\mathrm{cm^{-3}}$ quickly.
Figure~\ref{rho_r_1100002} shows gas density profiles at the
three output times. 
The density evolves self-similarly \citep{Larson1969,Penston1969},
but is perturbed tidally
at $R \simeq 4\;\mathrm{pc}$.

It is worth noting that, once the density exceeds 
$\simeq 3 \times 10^3\;\mathrm{cm}^{-3}$ with $T = 8000\;\mathrm{K}$, 
$\text{H}_2$ cooling never becomes efficient \citep{Inayoshi+2012}.
The regime is called ``zone of no return'', where 
collisional dissociation rapidly destroys
$\text{H}_2$ molecules.
In our calculation, the cloud enters the zone at
$t\simeq48.9\;\mathrm{Myr}$, after which $\text{H}_2$ cooling
cannot operate in a contracting gas.

It is important to estimate the mass of an SMS to be formed
in the gas cloud S1 and S2. As a rough estimate, we calculate
the Jeans mass and find that 
it is comparable to the enclosed gas mass 
at $M(<R) \sim 10^5~M_\odot$,
with a negligible contribution of DM.
Hence, the gas is self-bound and is going to collapse.
Figure~\ref{Mdot_1100002} shows the radial
distribution of the mass infall rate
$\dot{M} \equiv 4 \pi R^2 \rho(R) v_\text{in}(R)$ 
for S1 (red) and S2 (blue) when the central density reaches 
$2 \times 10^7\;\mathrm{cm}^{-3}$. Here, $v_\text{in}(R)$ 
is the radial infall velocity component.
We suppose that a protostar will accrete the surrounding gas
approximately at the instantaneous mass infall rate. 
The expected mass infall rates are 
$0.1 \sim 1 \;M_\odot\;\mathrm{yr^{-1}}$ at
$10^3~M_\odot \lesssim M \lesssim 10^5~M_\odot$ 
for S1 and S2. 
At $M \gtrsim 10^5~M_\odot$, however, 
the mass is outflowing from the cloud 
because the tidal force strips the gas 
in the halo outskirts.

The obtained $\dot{M}$ is very large, as expected in the DC model.
We infer the final stellar mass of the SMS 
without following the evolution in the accretion stage.
The stellar evolution calculations show that the stellar envelope 
inflates with such rapid mass accretion
\citep[e.g.,][]{OmukaiPalla2003,Hosokawa+2012}.
The stellar effective temperature is fixed nearly at 
$\simeq 5000 \;\mathrm{K}$ in the so-called supergiant stage. 
As a result, the emissivity of ionizing photons 
is greatly reduced \citep[e.g.,][]{Hosokawa+2013,Schleicher+13,Sakurai+2015}.
The radiative feedback will be too weak to halt the mass accretion, 
so that the star finally accretes almost all the infalling gas. 
The estimated mass from the infalling rate is $2 \times 10^5 \;M_\odot$ 
for S1 and $3\times 10^5 \;M_\odot$ for S2.

We follow the collapse of the DC clouds
until the central density reaches $ 10^{8}~\mathrm{cm}^{-3}$.
The angular momentum could arise as a barrier against the collapse, 
potentially decelerating the collapse to cause gravitational 
fragmentation. 
However, numerical simulations show that the clouds do not
have such high angular momentum because of 
the gravitational torque caused by the nonaxisymmetric 
structure of the host DM halo
\citep{Choi+2015, Shlosman+2016, Luo+2016}.
Once the collapse begins, vigorous fragmentation
does not occur until
an embryo protostar forms \citep[e.g.,]{Inayoshi+2014, Latif+2016}.
We have run two additional simulations to higher densities.
Our preliminary simulations also show that our DC clouds, S1 and S2, 
also continue to collapse until reaching $n \sim 10^{13}~ \mathrm{cm}^{-3}$
without fragmentation. 
The fragmentation will occur more often 
in the late stage of the protostellar accretion owing to the
gravitational instability of a circumstellar disk
\citep[e.g.,][]{Becerra+2015, Sakurai+2016}. 
To know whether the formation of a $10^5 \;M_\odot$ star is possible, 
we have to follow a long-term evolution with accretion.
This is beyond the scope of this paper and will be studied in our next work.

\begin{figure}[tb]
	\centering
	\includegraphics[width=8.3cm]{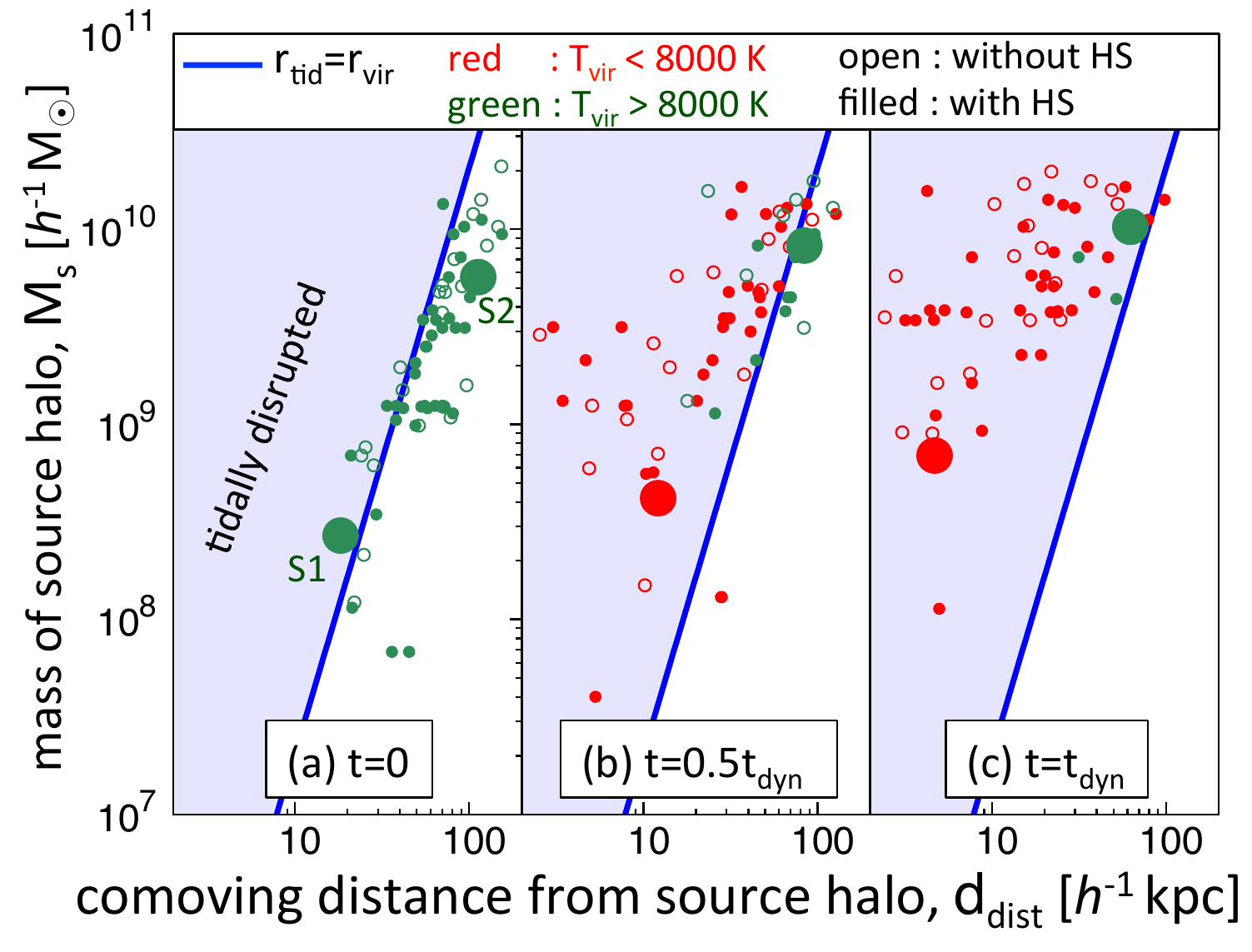}
	\caption{
Time evolution of the source halo mass 
$M_\text{s}$ and the comoving distance between the candidate and source
halos $d_\text{dist}$ for all 68 candidate halos.
Panels (a), (b), and (c) show the snapshots at the different epochs
$t=0$, $0.5$, and $1.0t_\text{dyn}$, respectively.
The candidate halo satisfies the DC criteria at $t=0$. 
The symbol colors represent the halos with $T_\text{vir} < 8000\;\mathrm{K}$ (red)
and $> 8000\;\mathrm{K}$ (green). 
The filled and open circles show whether the hydrodynamical simulations
(HS) are performed or not. 
The halos that host the collapsed clouds (S1 and S2) are shown
 with the larger circles.
In each panel, the shaded area denotes $R_\text{tid} < R_\text{vir}$,
i.e., the tidal force is strong enough to disrupt the halo envelope.
The solid line represents the critical points for $R_\text{tid} = R_\text{vir}$.   
}
	\label{rtid_to_smass}
\end{figure}

\begin{figure}[thb]
	\centering
	\includegraphics[width=7.7cm]{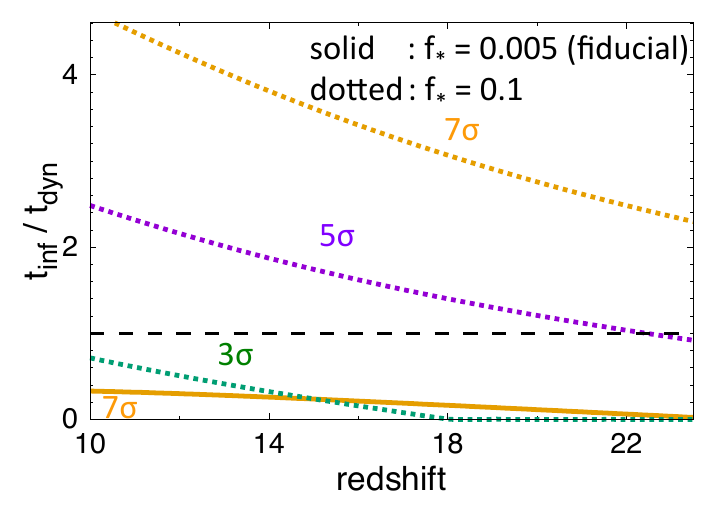}
	\caption{
	Infall time of the target halo 
	from $d_\text{LW}$ (eq. \ref{eq_d_LW}) to 
	$d_\text{tid}$ (eq. \ref{eq_d_tid}) normalized by the dynamical
	time at each redshift.
	The lines represent the evolution for the source halos
	that are formed from
	$3\sigma$ (green), $5\sigma$ (purple), and $7\sigma$ (orange)
	initial density peaks.
	The solid and dotted lines are for the fiducial and enhanced star formation efficiency,
	$f_*=0.005$ and $0.1$, respectively.
	The other two cases with $3\sigma$ and $5\sigma$ are not shown
	because $d_\text{LW}$ is always smaller than $d_\text{tid}$.
	The horizontal dashed line indicates $t_\text{inf} = t_\text{dyn}$.
	}
\label{t_infall}
\end{figure}

\section{Cloud Collapse under Strong Tidal Field} \label{sec_coll_tidal}
In this section, we quantitatively study the tidal effect
on the evolution of DC candidate halos.
Throughout this section, we refer to a candidate halo
as the ``target'', while a halo that is emitting
$\mathrm{H}_2$ dissociation radiation is referred to as the ``source''.

We first consider how strong the tidal force is
at the virial radius $R_{\rm vir}$ of the target halo,
which we compare with the tidal radius given in
eq. (15).
The tidal force by the source halo gets stronger
as the target halo approaches. 
Thus, the cloud (in the target halo) collapses within
an infalling timescale
after the target halo satisfies DC criteria.

\subsection{Evolution of Tidal Radius} \label{sec_tinf_sim}

We plot the time evolution of the source halo mass
and separation between the source and target halos
in Figure~\ref{rtid_to_smass}.
Each panel shows the snapshot at (a) $t=0$, (b) $0.5$, 
and (c) $1.0~t_\text{dyn}$ after the target halo satisfies the DC criteria.  
The shaded region indicates $R_\text{tid} < R_\text{vir}$, where the
outer part of the candidate halo is subject to the tidal force
from the source halo.  

In most of the cases, the virial temperature of the halo falls below 
$8000\;\mathrm{K}$ as they approach the source halo.
The target halo loses its mass by  
tidal disruption, and the gas cloud does not 
collapse. Clearly, the tidal disruption prevents
the clouds from collapsing.
The target halos fall in faster
than collapsing.
With our model of a UV-emitting galaxy,
the LW intensity at the position of the target halo
at distance $d_\text{dist}$ from the source halo is given by \citep{Iliev+2005b};
\begin{eqnarray} \label{J_est}
J_{21} = 63  \left ( \frac{M_\text{source}}{10^9\;M_\odot} \right ) &&\left ( \frac{f_\gamma}{2.0}  \right ) \nonumber \\
\left ( \frac{f_\text{J21}}{0.4}  \right )^{-1} && \left ( \frac{20\;\mathrm{Myr}}{t_\text{s}} \right ) \left ( \frac{1\;\mathrm{kpc}}{d_\text{dist}} \right )^2 ,
\end{eqnarray}
where $t_\text{s}$ is the mean lifetime of UV-emitting stars in the
source halo, 
$f_\text{J21}$ is the fraction of the LW intensity originating from the closest luminous halo to the total LW intensity,
$f_\gamma  $ is defined as $f_\gamma \equiv f_*
f_\text{esc} N_\text{i}$, where $f_*$ is the star formation efficiency,
which is the baryon fraction converted to stars,
$f_\text{esc}$ is the escape fraction of the UV photons, 
and $N_\text{i}$ is the number of the emitted photons 
per unit stellar mass throughout the mean lifetime $t_{\rm s}$.  
We take the fiducial value of $f_\gamma = 2$ \citep{Iliev+2007} and $f_\text{J21} = 0.4$.
The latter choice is motivated by the result of our semianalytical model (Section \ref{sec_SAM}).

Equation \eqref{J_est} gives the critical distance $d_\text{LW}$,
within which the target halo is irradiated by sufficiently strong LW 
radiation for DC. With standard normalizations,  
the critical distance is given by
\begin{eqnarray}
d_\text{LW} &=& 0.2 \; \mathrm{kpc}  \left ( \frac{M_\text{source}}{10^8\;M_\odot}  \right ) ^{1/2} \left ( \frac{f_\gamma}{2} \right ) ^{1/2}
 \nonumber
\\  &&  \left ( \frac{f_\text{J21}}{0.4}  \right )^{-1/2} 
\left ( \frac{20\;\mathrm{Myr}}{t_\text{s}} \right ) ^{1/2} \left ( \frac{100}{J_{21}^\text{crit}} \right ) ^{1/2}.  \label{eq_d_LW}
\end{eqnarray}
The distance at which the tidal radius is equal to the virial radius, 
$d_\text{tid}$, can be estimated by eq. \eqref{eq_tidal} as
\begin{equation} \label{eq_d_tid} 
d_\text{tid} = 0.524 \;h^{-1}\;\mathrm{kpc} \; \left ( \frac{M_\text{source}}{10^8\;h^{-1}\;M_\odot} \right ) ^{1/3} \left ( \frac{10}{1+z} \right ), 
\end{equation}
where $z$ is the redshift under consideration.

A cloud can collapse when it lies in $d_\text{tid} < d_\text{dist}  < d_\text{LW}$,
if the collapse is initiated at the virial radius.
Because the source halo also grows in mass, the ratio of $d_\text{LW}$ to
$d_\text{tid}$ becomes larger and the target halo might spend for a longer
time in $d_\text{tid} < d_\text{dist}  < d_\text{LW}$, because $d_\text{LW} \propto M_\text{source}^{1/2}$
while $d_\text{tid} \propto M_\text{source}^{1/3}$.

We evaluate the infalling time for the target halo moving from $d_\text{LW}$
to $d_\text{tid}$, considering the infalling velocity at
$d_\text{LW}$. Detailed derivations based on a spherical collapse model
are given in Appendix \ref{sec_Append}. 
The derived infall time $t_\text{inf}$ is determined by the source halo mass but
independent of the mass of the target halo.

Figure~\ref{t_infall} shows the infalling time $t_\text{inf}$ for the 
source halo formed in $3\sigma$ (green), $5\sigma$ (purple),
and $7\sigma$ (yellow) peaks of the initial density field. 
The solid line corresponds to the case with the fiducial parameters used 
in our semianalytical model.
We select $3\sigma$--$4\sigma$ regions,
so the candidate halos 
should be
disrupted by the tidal force from
the source halo before they collapse, aside from a few exceptions 
for which the major merger accelerates the collapse. 
Even for a $7\sigma$ halo, which is as rare as 
the observed high-$z$ QSOs ($\sim ~\mathrm{Gpc^{-3}}$),
$t_\text{inf}$ is much smaller than $t_\text{dyn}$.
The dashed lines show the infalling time of the target halo 
under the enhanced star formation efficiency $f_* = 0.1$,
much larger than the typical value $f_*=0.005$.
This is the highest value of $f_*$ allowed to explain
the observed galaxy luminosity functions at $z\sim 6$--$7$ \citep{Agarwal+2012}.
The infalling time $t_\text{inf}$ can exceed $t_\text{dyn}$ in higher-density peaks,
while they remain to be of an order of unity even for $7\sigma$ peaks
in the shown redshift range.
This fact implies that
the tidal force still has some impact on the collapse of the target halo
because $t_\text{inf}$ remains comparable to $t_\text{dyn}$.

The above discussions are only concerned with the tidal force at the virial
radius. When an infalling cloud begins collapsing near the virial radius
of the target halo,
it approaches the source halo 
so fast that the tidal force becomes strong and prevents the collapse.
Therefore, for a cloud to experience DC, gravitational instability needs to take place
well within the inner part of the target halo,
where the tidal effect from the source halo is weak.

\begin{figure}[htb]
	\centering
	\includegraphics[width=7.7cm]{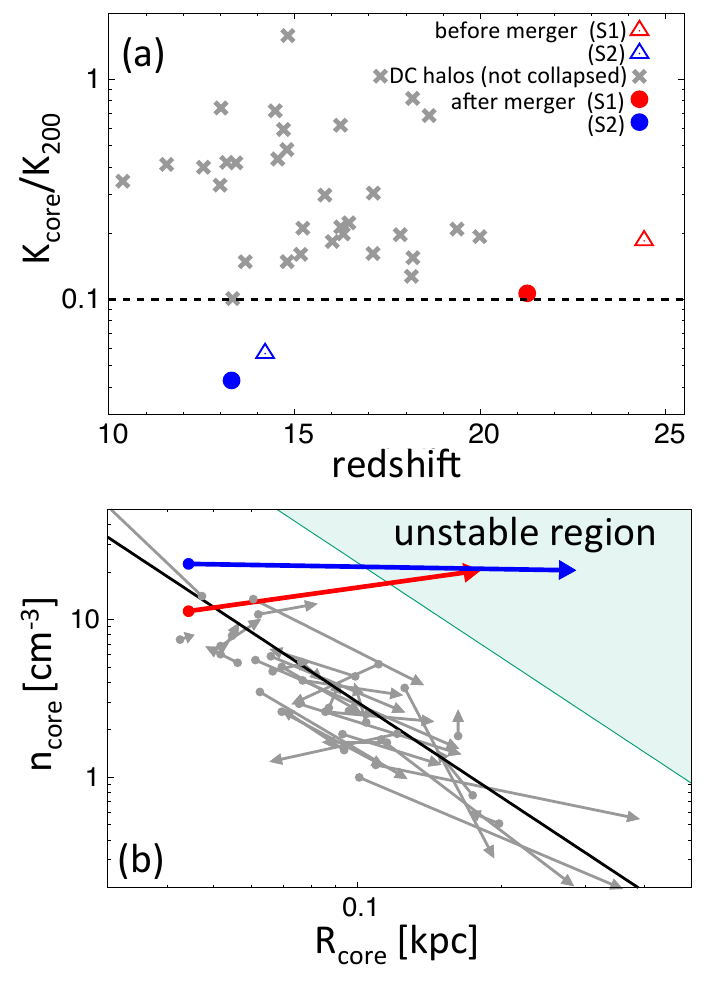}
	\caption{(a) Entropy of the cloud core normalized by $K_{200}$ (eq. \ref{eq_K200}). 
	The colored symbols denote the collapsed halos S1 (red) and S2 (blue),
	whereas the gray symbols show the uncollapsed clouds.
The open and filled symbols represent the entropy of
collapsed clouds before and after the major merger, respectively. 
The dashed line shows the mean value of the entropy by the universal profile.
(b) Evolution of the core radius and number density after
the host halo satisfies the DC criteria. 
The start (filled circle) and end points of each arrow represent 
the core properties when the halo meets the DC criteria and 
$1 t_\text{dyn}$ after that moment, respectively.
The black solid line shows the analytical core 
radius--density relation given by eq. \eqref{eq_core}.
The cloud becomes gravitationally unstable in the shaded
region, satisfying eq. (\ref{eq:nr_unstable}).
}
	\label{Entropy_core}
\end{figure}

\subsection{Onset of the Cloud Collapse} \label{sec_DC_collapse}

The DC (candidate) clouds evolve adiabatically, without significant
radiative cooling, until the gas temperature
reaches $8000\;\mathrm{K}$ because $\text{H}_2$ cooling is inefficient. 
It is known that a gas cloud contracting adiabatically 
develops a universal entropy profile, as studied 
in the context of the formation of galaxy clusters
\citep[e.g.,][]{Voit+2003,Voit+2005}. 
In spite of the enormously different mass scale, our simulated gas clouds
hosted by the DC candidate halos also show such 
universal features \citep[see also][]{Visbal+2014}.

The entropy of an ideal gas is defined 
by the thermodynamical quantities as
\begin{equation}
K = k_\text{B} T n_\text{b}^{-2/3},
\end{equation}
where $k_\text{B}$ is the Boltzmann constant, $T$ is the temperature,
and $n_\text{b}$ is the baryon number density. 
We can also define a ``halo entropy'' as
\begin{equation}
K_{200} = k_\text{B} T_\text{vir} ( 200\bar{n}_\text{b})^{-2/3}, \label{eq_K200}
\end{equation}
where $\bar{n}_\text{b}$ is $\Omega_\text{b}/\Omega_\text{m}$ times 
the mean matter density of the universe.
The entropy profile of the cloud consists of two parts, a constant
entropy core and an outer envelope with $K \propto (r/r_0)^{1.1}$. Under the
adiabatic collapse, 
the core radius and entropy are approximately given as
$R_\text{core} \sim 0.1R_\text{vir}$ and $K_\text{core} \sim
0.1K_{200}$, respectively \citep{Visbal+2014}.
We find that our DC clouds roughly follow this universal profile. 
Figure~\ref{Entropy_core}(a) shows $K_\text{core}$ of the candidate
halos just after they satisfy the DC criteria.
The halos have slightly larger entropies than $0.1K_{200}$
but show little redshift dependence.
It is interesting that the two successful cases, S1 (red) and S2 (blue),
have the smallest entropies after the major merger.
Figure~\ref{Entropy_core}(b) plots the evolution
of all the DC clouds in the $n_\text{core}$ - $R_\text{core}$
from the time when the halo meets the DC criteria (initial point)
to $t_\text{dyn}$ after that moment.
The evolution is depicted by the arrows in the figure.
We define $R_\text{core}$ as the radius at which the density is
half of the maximum density at the center.
If we approximately set $K_\text{core} = 0.1 K_{200}$ and $R_\text{core} = 0.1 R_\text{vir}$,
the core properties can be given as follows \citep{Inayoshi+2015b}:
\begin{eqnarray}
R_\text{core} &=& 0.1 R_\text{vir}, \nonumber \\
                    &=& 35 \;h^{-1}\;\mathrm{pc} \left ( \frac{T_\text{vir}}{10^4\;\mathrm{K}} \right )^{3/2} \left ( \frac{1+z}{16} \right )^{-3/2}, \\
n_\text{core} &=& \left ( \frac{K_\text{core}}{K_{200}} \right ) ^{-3/2} \;\bar{n}_\text{b} = 22\;\mathrm{cm^{-3}} \left ( \frac{1+z}{16} \right )^3.
\end{eqnarray}
By assuming $T_\text{vir} = 8000\;\mathrm{K}$, we can get an analytical estimate
\begin{equation}
n_\text{core} = 0.03\;\mathrm{cm^{-3}} \left (\frac{R_\text{core}}{1\;\mathrm{kpc}} \right )^{-2}. \label{eq_core}
\end{equation}
The estimates are in good agreement with the core properties
of our sample (Figure~\ref{Entropy_core}b).

For a gas cloud to collapse, 
the core mass should be larger than the Bonnor--Ebert mass (eq. \ref{m_BE}).
The core mass can be estimated by
$\sim R_\text{core}^3 \rho_\text{core}$,
and thus the gravitational instability condition can be written as
\begin{equation}
\label{eq:nr_unstable}
n_\text{core} > 0.23 \;\mathrm{cm^{-3}} \left ( \frac{R_\text{core}}{1\;\mathrm{kpc}} \right )^{-2},
\end{equation}
which is shown by the shaded region in Figure~\ref{Entropy_core} (b). 
Note that this critical density is an order of magnitude higher than the
actual core densities. Figure~\ref{Entropy_core}(b) shows that
most of the candidate halos, except for the two successful cases, lie in the
stable region, even one dynamical time after the gas
starts cooling. Clearly, the core density needs to remain large to trigger DC.

We have seen that, in the cases of S1 and S2, the gas cloud collapse 
is driven by  major mergers. 
The high-speed mergers
with relative velocity of $v_\text{rel} \sim 2c_\text{s}$
generate shocks and increase the gas entropy. 
As shown in Figure~\ref{Entropy_core}(b),
the mergers increase not only the core density but also the core size
by an order of magnitude.
The S1 and S2 gas clouds move into the gravitationally unstable region
depicted in the figure.

\subsection*{}
Here we summarize the discussion in this section.
An adiabatically contracting gas cloud has
a universal entropy profile composed of a core and an envelope.
For the DC gas clouds in our simulation,
the core does not contain a large enough
mass to gravitationally collapse. 
The core would gain further mass through accreting the surrounding 
gas, but the host halo itself is tidally disrupted quickly within one dynamical time.
Gravitational collapse is often completely halted. 
The only ``successful'' path found in this paper is the sudden increase of the core mass
by major merger events. We have found that the successful DC systems
in our samples indeed experience major mergers.

\section{Discussion}

\begin{figure}[!t]
		\centering
		\includegraphics[width=8.1cm]{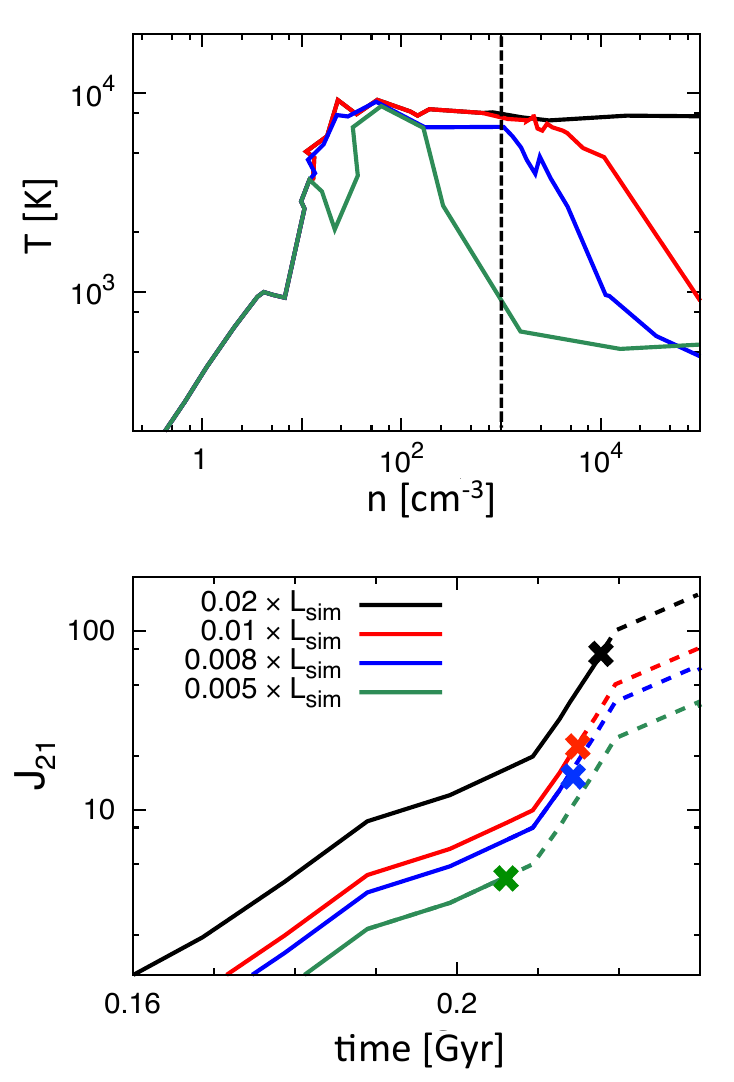}
\caption{(a) Thermal evolution of the collapsing cloud S1
with the smaller LW luminosities of the light source halo: 
0.5\% (green), 0.8\% (blue), 1\% (red), and 2\% (black) 
of the original value. 
(b) Time evolution of $J_{21}$ 
for the same cases.
The crosses indicate the epochs when the cloud density 
has reached $n_\text{LTE}=10^3\;\mathrm{cm}^{-3}$. 
The dashed line shows the evolution of LW intensity for $n>n_\text{LTE}$.}
\label{rho_T_Norm}
\end{figure}

\subsection{Critical LW Intensity for DC} \label{sec_Jcrit}

We have adopted a constant value of $J_{21}^\text{crit}=100$
as our fiducial critical value for Pop II sources \citep{Shang+2010}. 
We have also shown that the LW intensity at a cloud's
position varies with time
as it approaches the light source halo.
It is important to reexamine in detail the critical intensity $J_{21}^{\rm crit}$
for DC.

Whether or not a gas cloud evolves along the 
atomic-cooling track is determined by the LW intensity 
when the gas density reaches
$n_\text{LTE} \sim 10^3 \;\mathrm{cm^{-3}}$,  
the critical density for the $\text{H}_2$ rovibrational transitions
\citep[e.g.,][]{Omukai2001,Shang+2010}.
In order to study the effect of a time-dependent LW radiation field,
we perform several test 
simulations based on our sample S1.

We record the time evolution of the LW luminosity of the
light source halo $L_\text{sim}(t)$ in the original case S1. 
We then follow the collapse of the same gas cloud,
reducing the source LW luminosity to  
0.5\%, 0.8\%, 1\%, and 2\% of the original $L_\text{sim}(t)$.
Note that, for all these cases, the LW intensities
at the virialization are much lower than the 
normal critical value $J_{21}^{\rm crit} \sim 100$.
Figure~\ref{rho_T_Norm}(a) shows the evolution for the test cases.
We see that even 2\% of $L_\text{sim}(t)$ can lead to DC, 
whereas $\text{H}_2$ formation and cooling operate with 1\% of $L_\text{sim}(t)$.

Figure~\ref{rho_T_Norm}(b) shows the time evolution of the
LW intensity at the cloud core. 
The crosses indicate the epoch when the core density
has reached $n_\text{LTE} = 10^3\;\mathrm{cm^{-3}}$. 
The critical LW intensity at $n \sim n_\text{LTE}$
is between $20$ -- $80$, 
which is comparable to $J_{21}^\text{crit}$ obtained by \cite{Shang+2010}.
We thus suggest that the critical LW intensity for 
DC should be evaluated at $n \sim n_\text{LTE}$, 
not at the time of virialization. 
The figure also shows that a slight reduction of the source
luminosity results in a dramatically lower value of
$J_{21,n_\text{LTE}}$.
With 0.5\% of $L_\text{sim}(t)$, for instance, $J_{21,n_\text{LTE}}$ 
is lower by an order of magnitude than that with 2\% of
$L_\text{sim}(t)$, though the luminosity differs only by one-fourth. 
This is because, with 0.5\% of $L_\text{sim}(t)$, the collapse
advances earlier via efficient H$_2$ cooling when the cloud is
located at a distant place
from the source halo. 
Obviously, following both the cloud collapse in and around
the virial radius and the
halo assembly is 
necessary to derive the LW intensity at the critical density
$n_\text{LTE}$.

Overall, the above results suggest that setting $J_{21}^{\rm crit} \sim 100$ 
at the virialization is too strict for DC.
In general, it takes a few dynamical times from the virialization 
until the density reaches $n \sim n_\text{LTE}$. 
During this period, $J_{21}$ increases by an order of magnitude
as the cloud approaches the light source halo. 
Therefore, moderate LW intensity at the epoch of the virialization,
probably much lower than previously thought, 
is sufficient to induce DC. 

This opens a possibility that a large number 
of DC halos may exist in the universe. 
For the DC halos in a realistic cosmological context, however,
the exact value of $J_{21}^\text{crit}$ at virialization
actually depends on the subsequent evolution, 
which can be followed only by hydrodynamical simulations.
Ultimately, we will need a number of hydrodynamics simulations with a large volume,
with which we can find possibly more halos under the moderate LW fields
$J_{21} \sim 10$.
It is also necessary to follow the cloud collapse until
the density reaches $n \sim 10^3\;\mathrm{cm^{-3}}$ for each case.

\begin{figure}[tb]
		\centering
		\includegraphics[width=8.05cm]{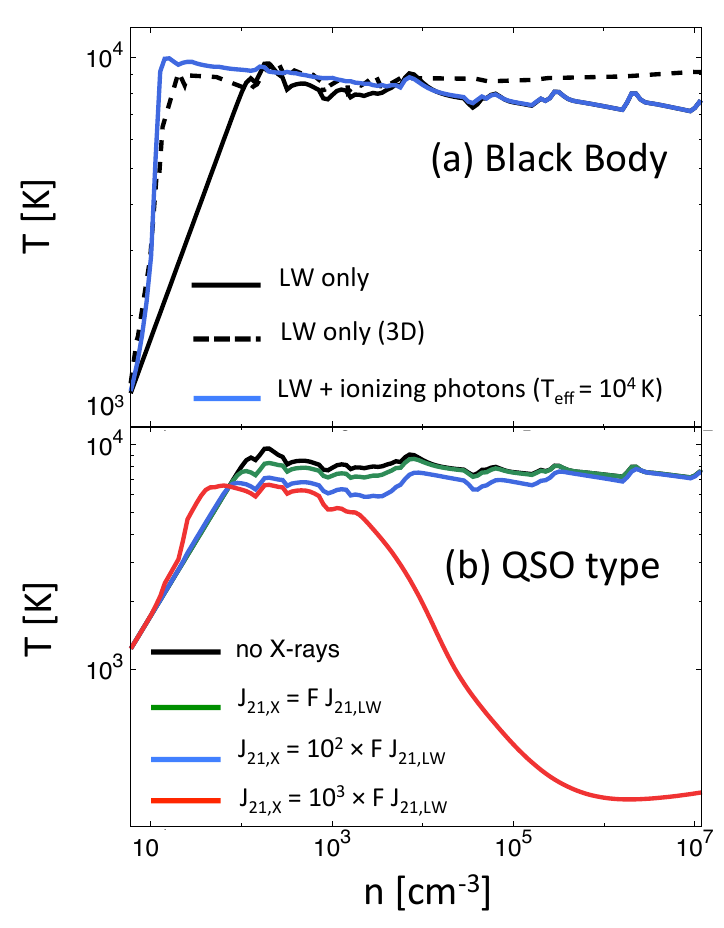}
		\caption{(a) Evolution of the temperature as the density rises with and without ionizing radiation. 
		The black line represents the case without UV photons, and the blue line represents the evolution under the black body radiation with 
		$T_\text{eff} = 10^4\;\mathrm{K}$. 
		The dashed line corresponds to the 3D simulation (S1) ,while the solid lines are the results of the one-zone calculation.
		In the presented one-zone models, we use the time evolution of 
		the density taken from the 3D simulation (see eq. \ref{eq_rho_evo} ).
		The amplitude of radiation is the same as the normalization of the intensity at the LW bands, $J_{21,\text{LW}}$.
		(b) Evolution of the temperature as the density rises with and without X-rays of the QSO-type spectra, 
		$J_{21} = J_{21,\text{X}} (\nu/\nu_0)^{-\alpha}$. 
		The black line shows the evolution without the X-rays, while the green, blue, and red lines show the evolution with 
		$J_{21,\text{X}} = 1$, $10^2$, and $10^3 \times F J_{21,\text{LW}}$, respectively.
		}
		\label{rho_T_Xion}
\end{figure}

\subsection{Effects of Ionization Photons and X-Ray} \label{sec_ionization_photons}
We have only considered UV photons with energies below $13.6\;\mathrm{eV}$ because
the mean free path of ionizing photons is much smaller than that of
the LW photons. If a collapsed cloud approaches the vicinity
of the light source halo, the cloud might be affected directly
by ionizing photons. 

The ionizing radiation enhances $\text{H}_2$
cooling, because the electrons catalyze the formation
of $\text{H}_2$ via the $\text{H}^-$ reaction path. 
The increase of the ionization degree accelerates the formation
of $\text{H}_2$ and thus may prevent DC.
In addition, X-ray photons produced by the
light source have large mean free paths, which can have
an impact on the evolution of the central core of
the cloud \citep{Inayoshi+2011,Inayoshi+2015}. 

It is clearly important to address the impact of the ionizing photons and X-rays
on the thermal evolution of DC clouds.
To this end, we employ one-zone calculations and follow the evolution of the S1 cloud
by including the effects of the additional radiation. 
We directly use the density evolution in our simulation
because the cloud actually collapses over a much longer timescale than $t_\text{ff}$,
while we assume that the collapse proceeds at a rate of $t_\text{ff}$
when $\text{H}_2$ cooling is efficient.
The density evolution can be described as 
\begin{eqnarray}
\frac{\mathrm{d}\rho}{\mathrm{d}t} = \left \{ 
\begin{array}{lll}
\rho/t_\text{ff} \;\;\;&& (\text{when}\;\text{H}_2\;\text{cooling is efficient}),  \\
\dot{\rho} _\text{sim} (t) \;\;\; &&\text{(otherwise)}, 
\end{array}
\right .
\label{eq_rho_evo}
\end{eqnarray}
where $\dot{\rho} _\text{sim} (t)$ is the time derivative of the density 
of the cloud core in the simulation.
Nonequilibrium chemistry and the energy equation are solved in the same manner
as in our 3D simulation described in Section \ref{sec_hydro}.
The column density $N_i$ of each species $i$ is given by  
\begin{equation}
N_i = 0.5 \;\lambda_\text{J} n y_i,
\end{equation}
where $\lambda_\text{J}$, $n$, and $y_i$ are the Jeans length, the gas density, 
and the number fraction of the species $i$ of the cloud core, respectively. 
The self-shielding against the ionizing photons and the secondary ionization
are also considered, following \cite{Wolfire+1995}.
The LW intensity is assumed to be the same as the
intensity in the 3D simulation (S1).

We run the one-zone calculations for two different types of spectra:
a black body spectrum with  $T_\text{eff} = 10^4 \;\mathrm{K}$,
and soft X-ray radiation with a power-law
$J_{21} \propto J_{21, \text{X}} \left (\nu/\nu_0\right )^{-\alpha}$, 
with $h\nu_0 = 1\;\mathrm{keV}$ and $\alpha = 1.8$.
The conditions assumed above are, respectively,
that the DC cloud is located within the Str\"omgren
sphere around its light source halo and
that the cloud is irradiated
by X-ray radiation from the source. 
The spectrum resembles that of stars and QSOs, respectively.
As for the QSO-type spectrum, we set the minimum and the maximum
energy as $h\nu_\text{min} = 1 \;\mathrm{keV}$ and $h\nu_\text{max} = 10 \;\mathrm{keV}$.

The X-ray observations of local starburst galaxies show a correlation
between the X-ray luminosities and SFR \citep{Glover+2003,Grimm+2003}:
\begin{equation}
L_\text{X} = 6.7\times 10^{39} \left (\frac{\text{SFR}}{M_\odot~ \mathrm{yr}^{-1} } \right ) \mathrm{erg} \; \mathrm{s}^{-1}.
\end{equation}
There is also a correlation between the LW luminosities and SFR \citep{Inoue2011}:
\begin{equation}
L_\text{LW} = 1.14\times10^{43}  \left (\frac{\text{SFR}}{M_\odot~ \mathrm{yr}^{-1} } \right ) \mathrm{erg} \; \mathrm{s}^{-1}.
\end{equation}
By eliminating SFR in the above two observational correlations, 
we obtain the relation between the intensity in the LW bands 
$J_\text{21,LW}$ and $J_\text{21,X}$ as follows \citep{Inayoshi+2015}:
\begin{equation}
J_\text{21,X} = 5.7 \times 10^{-6} \;J_\text{21,LW} \equiv F J_\text{21,LW}.
\end{equation}

Figure~\ref{rho_T_Xion}(a) shows the evolution with
the stellar radiation case.
We normalize the ionization radiation intensity
at the Lyman limit, $J_\text{UV}$, to be $J_\text{21}$.
Due to ionization heating, the temperature rises rapidly 
to $T \sim 10^4 \; \mathrm{K}$. 
The subsequent collapse proceeds almost isothermally by
efficient Lyman-$\alpha$ cooling
The cloud becomes optically thick to the ionization radiation
when $n \sim 10^3 \;\mathrm{cm^{-3}}$,
and the thermal evolution converges to that of the case
without ionizing radiation.
This is consistent with the 1D radiation hydrodynamics
calculation of \cite{Kitayama+2001}.

Figure~\ref{rho_T_Xion}(b) shows the evolution 
with the QSO-type radiation. 
For comparison, the evolution with $10^2$ and
$10^3$ times stronger X-ray intensities is
also calculated. In the case with $10^3$ times stronger X-ray intensity,
DC does not occur.
This case with a very strong X-ray intensity
is an extreme example, and we do not expect that such conditions
are realized in the early universe.
Overall, neither ionizing radiation nor X-rays have a significant
impact on the thermal
evolution of the DC cloud of S1
\citep[note a similar conclusion by][for the former case]{Inayoshi+2015}.
\footnote{
The temperature fluctuation that appeared in the one-zone calculation (Fig.~\ref{rho_T_Xion})
is caused by using the density evolution from the hydrodynamical simulation.
The fluctuation amplitudes are so small that they do not affect the
overall evolution of the temperature.
}

\subsection{Effect of Metal Enrichment} \label{sec_metal_enrichment}

Our model includes metal 
enrichment within progenitor halos, 
but we do not consider ``external'' metal enrichment 
by nearby halos. 
We have shown that
the DC halos in our simulations
approach very close ($\sim 100 \;\mathrm{pc}$) to
the light source halos.
The distance is well inside the virial radius of the source halo 
$\simeq 700\;\mathrm{pc}$, and thus the DC halos could be polluted by 
metals dispersed from the star-forming (light source) halo by,
e.g., galactic winds.

We can estimate how far the metals will be dispersed around the 
star-forming halo as follows. 
The dynamics of an SN-driven bubble in the expanding universe
is approximately described by the Sedov--Taylor-type self-similar 
solution \citep{Voit1996}. 
The time evolution of the position $R$ of a shell 
around the bubble is described by
\begin{eqnarray} \label{eq_sn_st}
R = 23.6~\mathrm{ckpc} &&
\left [
\left ( \frac{f_*}{0.005} \right ) \left( \frac{f_\text{b}}{0.16} \right ) \left( \frac{f_\text{esc}}{0.1} \right ) 
\left( \frac{M_\text{halo}}{10^8~M_\odot} \right ) 
\right ]^{1/5} \nonumber \\
&&\left(\frac{21}{1+z} \frac{\hat{t}}{10^{10}~\mathrm{yr}} \right )^{2/5} 
\left ( \frac{\alpha}{0.5} \right ),
\end{eqnarray} 
where $\mathrm{d}\hat{t} = (1+z)^2\mathrm{d}t$, $t$ is the elapsed time after the SN 
explosion, and $f_\text{esc}$ is the fraction of the energy injected 
into the wind to the total SN explosion energy. 
We have used the conversion efficiency between the SN explosion energy
and stellar mass assuming the Salpeter IMF.
The factor $\alpha$ represents uncertainty originating from effects
such as the gravitational force from the host halo, external pressure, etc.
More detailed numerical calculations suggest that this factor is around $0.5$
\citep{Barkana+2001,Kitayama+2005}, which is adopted in equation~\eqref{eq_sn_st}.
Since the above size is smaller than or comparable to the virial radius of the light source halo,
we only consider potential effects of the metal enrichment
after the DC cloud plunges into the light source halo.

In our two cases S1 and S2, the density of the DC cloud reaches
$\sim 100\;\mathrm{cm^{-3}}$ when it passes through the
virial radius of the light source halo.
\cite{Cen&Riquelme2008} point out that, 
once such a dense core is formed, 
metal mixing into the core takes a much
longer time than the local dynamical time.
\cite{Smith+2015} also show that the densest part of a cloud
collapses before it is significantly enriched with
metals transported by mixing.
It would be interesting to explore cases where
the external metal enrichment  
from the light source halo prevents DC under peculiar conditions.

\subsection{Observational Signatures of DC Halos} \label{sec_CR7}

Recently, a strong helium-line-emitting galaxy without signatures
of heavy elements has been discovered by \cite{Sobral+2015}.
Several authors propose
that the galaxy, called CR7, may be powered by an intermediate-mass
BH hosted by a primordial halo
as expected in the DC model \citep[e.g.,][]{Pallottini+2015, Agarwal+2016,
Hartwig+2016}. CR7 actually consists of three components: a 
 helium line emitter without metals, and two small galaxies with spatial
 separations of $\simeq 5$~kpc. It is interesting, even encouraging,
 that our successful DC systems, with a DC halo and nearby light source halo(s),
 resemble the observed structure of CR7
\citep[e.g.,][]{Pallottini+2015}.

For our S1 and S2, however, the separations between
the DC and light source halos are much smaller than 5~kpc. 
Such small separations are a general feature of the DC candidate halos in
our simulations. Also, the resulting short separations suggest
strong tidal force that affects whether the main cloud collapses
(Section \ref{sec_coll_tidal}). 
It will be interesting to examine whether the
structure of CR7 is really realized in cosmological hydrodynamical simulations.

\section{Conclusion}

We have performed SPH/$N$-body simulations to
follow the evolution of gas clouds in early DM halos.
We find two cases out of 42 samples, where
the cloud collapses gravitationally
and the gas condenses to densities of 
$\sim 10^8~{\rm cm}^{-3}$. The two cases are clear examples
of DC realized in the cosmological setup.
The other 40 cases (halos) are identified as potential DC systems,
but the clouds do not collapse 
because of the dynamical interaction between the host
halo and the light source halo. 
In many cases, this is inevitable because strong LW radiation
from a nearby massive halo is a necessary condition for the DC.
Also, ram pressure stripping disturbs the cloud collapse moving
in a dense environment.

In the two successful cases (S1, S2), we expect the gas cloud
to yield a very massive star by mass accretion.
The estimated accretion rates from the radial infall rates 
are $\dot{M} \sim 0.1 - 1 \;M_\odot \; \mathrm{yr^{-1}}$
for $M (<R) \lesssim 10^5~M_\odot$.
Interestingly, the total mass of the infalling gas is reduced
by an environmental effect. The outer part of the cloud core is
subject to tidal disruption.
Clearly, the tidal force exerted by nearby halos plays an important role,
by disrupting the gas cloud itself in the failed cases,
and by limiting the available gas mass for star formation
in the successful cases.
 
We have also revisited the critical LW intensity required 
for the DC, considering the environmental effects included 
in our simulations. 
DC actually occurs even with 2\% of the
original LW intensity realized in our cosmological simulation.
The critical intensity for DC 
should be determined at the density of 
$n_\text{LTE} \sim 10^3\;\mathrm{cm^{-3}}$,
not at the virialization of atomic-cooling halos 
as has been often assumed in previous studies.
The LW intensity is highly variable in both time and space,
and thus it is necessary
to follow the evolution of a DC system using 
three-dimensional simulations in order to determine
robustly whether or not DC actually occurs.

Finally, we argue that the number density of the DC events
is significantly smaller than the estimates in the previous
studies. The expected number density of DC gas clouds,
two in a cube of 20 $h^{-1}~$Mpc on a side, if we naively take
our simulation result, 
is still larger than that of the observed high-$z$ quasars. 
While the DC model provides a viable mechanism for the formation
of `seed' BHs, we still appear to be a long way from
fully understanding the formation of SMBHs in the early universe.

\section *{Acknoledgement}
We thank K. Omukai, K. Sugimura, K. Inayoshi, G. Chiaki, Y. Sakurai, 
H. Susa, M. Umemura,
J. Wise, M. Latif ,T. Hartwig, M. Habouzit,
and M. Volonteri for fruitful discussions and comments.
This work was financially supported by Advanced Leading Graduate Course for Photon Science program
and by the Grants-in-Aid for Basic Research
by the Ministry of Education, Science and Culture of Japan
(25800102, 15H00776: T.H., 25287050: N.Y.) and by Grant-in-Aid for JSPS Fellows
(S.H.).
The numerical simulations were carried out on XC30 
at the Center for Computational Astrophysics (CfCA) of 
the National Astronomical Observatory of Japan.
We use the SPH visualization tool SPLASH \citep{SPLASH}
in Figs~\ref{halo_1701848}, \ref{halo_1200071}, \ref{halo_1100002}, and \ref{collapse_1100002}.

\bibliography{biblio2}

\appendix
\section{Particle de-refinement prescription} \label{sec_dezoom}

\begin{figure}[htb]
	\centering
		\includegraphics[width=8.0cm]{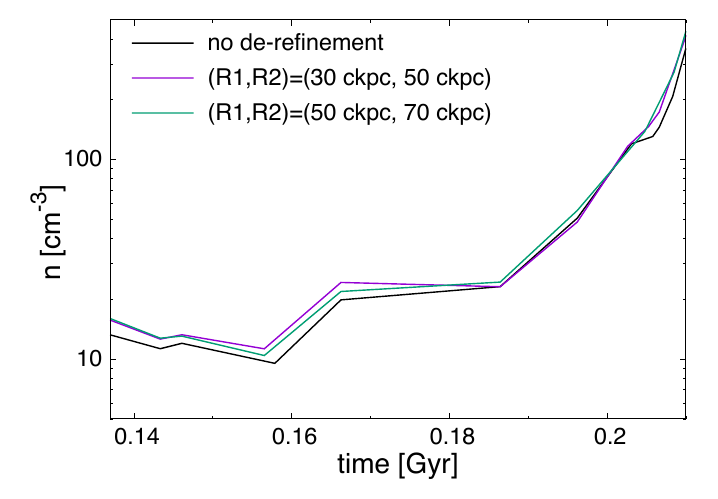}
		\caption{
Effects of varying the particle de-refinement criteria for 
the density evolution for S1 halo.
Each line shows the evolution without de-refinement (black) and
with de-refinement using $(R_1, R_2) = (30 \;\mathrm{ckpc}, 
50 \;\mathrm{ckpc})$ (purple) and 
$(R_1, R_2) = (50 \;\mathrm{ckpc}, 70 \;\mathrm{ckpc})$ (green).
}
		\label{dezoom}
\end{figure}

To speed up the hydrodynamical calculation, we develop a de-refinement 
method that combines and subtracts gas particles in the simulation.
We combine particles by the following steps:
\begin{itemize}
\item the region in which particles are combined is specified;
\item the particles are sorted along the Peano--Hilbert curve; and
\item particles arranged in the order are grouped and combined 
with every $N_\text{comb}$ particles.
\end{itemize}

We combine every eight particles into one particle to de-refine them 
so that the energy and momentum are conserved.
As control parameters, we set two distances from the center 
of the DC halo $R_1$ and $R_2$, where $R_1 < R_2$.
We carry out the de-refinement operation once for particles 
in $R_1 < R < R_2$ and twice for particles in $R_2 < R$, where  
$R$ is the distance between the particle and the DC halo center.

We perform test calculations for the S1 halo, for which the de-refinement
is done for the snapshot at $z = 24.6$, which corresponds to the cosmic age of $0.132$~Gyr. Figure~\ref{dezoom} shows 
the density evolution of the S1 halo with varying the 
de-refinement parameters $R_1$ and $R_2$. 
The difference caused by the parameter choice is only within a factor of two. 
Nonetheless, the computation time with 
$(R_1, R_2) = (30 \;\mathrm{ckpc}, 50 \;\mathrm{ckpc})$
is smaller by an order of magnitude than that with
$(R_1, R_2) = (50 \;\mathrm{ckpc}, 70 \;\mathrm{ckpc})$.

\def\thesection{\Alph{section}}
\section{The infalling velocities of DC candidate halos} \label{sec_Append}

We derive the halo infalling velocity $v_\text{inf}$ in the following steps:
(1) we first assess the linear overdensity $\delta_\text{in}$ and the mass $M_\text{in}$
within a distance $d_\text{LW}$, and (2) we calculate the turnaround 
radius $R_\text{turn}$ of the mass shell at $d_\text{LW}$ 
by the spherical collapse model. 
Here, the turnaround radius is defined by the radius at which the expansion of the mass shell
turns to the contraction.
The infalling velocity of the shell whose radius is $R$ is 
defined as $v_\text{inf} = \sqrt{2\mathrm{G}M(1/R - 1/R_\text{turn} )}$, 
where $M=M(R)$ is the enclosed mass within $R$.
The infalling time $t_\text{inf}$, the timescale 
over which the shell collapses from $R_1$ to $R_2$, is
\begin{equation}  
t_\text{inf} = \int^{R_1}_{R_2} \frac{\mathrm{d}R}{v} = 
\int^{R_1}_{R_2} \frac{\mathrm{d}R}{\sqrt{2GM(1/R - 1/R_\text{turn}) }}. 
\label{eq_t_inf}
\end{equation}
In what follows, we derive $M$ and $R_\text{turn}$ as functions of 
$R$, which enables us to perform the above integration.
Since we only consider the high-$z$ universe at $z > 8$, which is the matter-dominant
epoch, we assume the Einstein-de Sitter universe for simplicity below.

\subsection{Linear Overdensity within $R$} \label{lod_R}

The enclosed mass $M$ and the linear overdensity $\delta$ within the shell radius $R$ 
can be calculated by the spherical collapse model. 
Here, $\delta \equiv \rho / \bar{\rho} - 1$, where $\rho$ is the density and $\bar{\rho}$ is the mean density of the universe, which is equal to $\Omega_\text{m} \rho_\text{crit} (1+z)^3$. 
In the Einstein-de Sitter universe, the dynamics of the shell $R = R(t)$ 
is parameterized by $\theta$ as follows:
\begin{eqnarray}
\frac{R}{R_\text{l}(t)} &=& \frac{3}{10 \;\delta(t)}  (1-\cos \theta), \label{scm_radius} \\
\delta(t)                   &=& \frac{3}{5} \left [ \frac{3}{4}  (\theta-\sin \theta) \right ]^{2/3}  \label{scm_delta} ,
\end{eqnarray}
where $R_\text{l}(t) = [3M/(4\pi\bar{\rho}(t))]^{1/3}$. 
At the fixed $t$, two shells that have $\delta_i$ and $\theta_i$, where $i=1$ and $2$, have a relation as follows:
\begin{equation}
\frac{\delta_1}{ \delta_2} = \left ( \frac{\theta_1 - \sin \theta_1}{\theta_2 - \sin \theta_2} \right )^{2/3} .
\end{equation}

The completely collapsed shell with $r=0$ is characterized by $\theta = 2\pi$ 
and $\delta = \delta_c \equiv 1.69$.
Assuming that the source halo has just reached $\theta_2 = 2\pi$ and 
corresponds to $\delta_2 = \delta_c$, then
\begin{equation} \label{ratio_delta_2pi}
\delta_1 = \delta_c \left ( \frac{\theta_1 - \sin \theta_1}{2\pi} \right )^{2/3}.
\end{equation}
With eqs. \eqref{scm_radius} and \eqref{ratio_delta_2pi}, the shell radius 
is determined by the enclosed mass and the linear overdensity as $R = R(M, \delta)$.

\subsection{Enclosed Mass and Turnaround Radius}
We evaluate the enclosed mass $M$ within the shell radius $R$ following \cite{Barkana2004}, which
is based on the extended Press--Schechter (EPS) theory in the Fourier space \citep{Bond+1991}
and the Press--Schechter (PS) theory in the real space.
EPS theory can handle the contribution of the negative overdensity to the collapsed halo mass,
while 
the negative overdensity never contributes in the PS theory and thus 
requires ad hoc treatment.
In the EPS theory, it is convenient to evaluate the density field in the Fourier space.
Instead of the wavenumber $k$, the variance 
$S_k \equiv 1/(2\pi^2) \int_0^k \mathrm{d}k' k'^2 P(k')$
is used to represent the scale under consideration, where $P(k)$ is the 
power spectrum of the initial density fluctuation. 

EPS theory provides the probability of the density lying between 
$\delta$ and $\delta + \mathrm{d}\delta$,
$Q(\nu,\delta,S_k) \mathrm{d}\delta$, 
where $\nu$ is the critical overdensity for the collapsing halo.
The cumulative mass function is obtained by integrating $Q$ from $\nu$ to $\infty$, 
while its differentiation gives the differential mass function $f(\nu, S_k) \mathrm{d}S_k$. 
$Q$ and $f$ can be written as follows:
\begin{align}
&Q (\nu,\delta,S_k) = \frac{1}{\sqrt{2\pi S_k}} \nonumber \\
&\left [ \exp \left (  -\frac{\delta^2}{2 S_k} \right ) - \exp \left (  -\frac{(2\nu - \delta)^2}{2 S_k} \right ) \right ] , 
\end{align}
\begin{eqnarray}
f (\nu, S_k) &=& \frac{\partial}{\partial S_k} \left [1 - \int_{-\infty}^\nu \mathrm{d}\delta Q (\nu,\delta,S_k) \right ]
\nonumber \\ &=& \frac{\nu}{\sqrt{2\pi} S_k^{3/2}} \exp \left ( -\frac{\nu^2}{2S_k} \right ).
\end{eqnarray}
The probability distribution $(\delta, S_k)$ around $(\delta_\text{c}, S_{k,M})$ is
given by
\begin{align}
P (\delta | \delta_\text{c}) = Q (\delta_\text{c},\delta,S_{k,M}) \frac{f(\delta_\text{c}-\delta,S_{k,M} - S_k)}{f(\delta_\text{c},S_{k,M})}.
\end{align}
Therefore, the mean density profile $\langle \delta(r) \rangle$ around $(\delta_\text{c}, S_{k,M})$ becomes
\begin{align} \label{eq_delta_r}
&\frac{\langle \delta(r) \rangle}{\delta_\text{c}} 
= 1 - \left (  1 - \alpha + \frac{\alpha}{\beta} \right ) \nonumber \\
&\text{erf} \left [ \sqrt{\frac{\beta (1-\alpha)}{2\alpha}}\right ]
-\sqrt{\frac{2\alpha (1-\alpha)}{\pi \beta}} \exp \left [ -\frac{\beta (1 - \alpha)}{2\alpha}\right ], 
\end{align}
where $\alpha \equiv S_k/S_{k,M}$ and $\beta \equiv \delta_\text{c}^2 / S_{k,M}$.

It is pointed out by \cite{Barkana2004} that
in the rare halo limit where $\beta \sim 0$, the PS theory describes the density distribution
more accurately than the EPS theory. 
In the PS theory, the mean density profile $\langle \delta (r) \rangle$ is given by
\begin{equation}
\frac{\langle \delta(r) \rangle}{\delta_\text{c}} = \frac{\xi_r(r_M,r)}{\sigma^2(r_M)}, 
\end{equation}
where $\xi_r$ is the two-point correlation function.
To compromise the EPS and PS theories, he adopts 
\begin{equation} \label{eq_ab_PS_EPS}
\alpha = \frac{\xi_r}{\sigma^2(r_M)},\;\; \beta = \frac{\nu^2 \alpha (1-\alpha)}{\sigma^2(r) - \alpha \xi_r(r_M, r)},
\end{equation}
instead of the values in the previous discussion.

Once the enclosed mass $M$ is given, then we can derive the averaged 
density within the shell, $\delta = \delta(M)$ from eqs. 
\eqref{eq_delta_r} and \eqref{eq_ab_PS_EPS}. 
According to the discussion in Appendix \ref{lod_R}, the shell radius $R$ 
and enclosed mass are related by $R = R(M, \delta(M))$. 
Therefore, for a given radius $R$, $M$, and $\delta$ within $R$ are 
obtained by solving the equation $R = R(M, \delta(M))$.

With the obtained overdensity $\delta$ within the radius $R$, 
we can relate $R$ and the turnaround radius $R_\text{turn}$
corresponding to $\theta = \pi$ and $\delta = 1.06$ using 
eqs. \eqref{scm_radius} and \eqref{scm_delta} \citep{MoBoschWhite2010},  
\begin{equation}
R_\text{turn} = \frac{ 2 }{ 1 - \cos \theta_1} R,
\end{equation}
where $\theta_1$ is the parameter representing the 
considered shell given by eq. \eqref{scm_delta}.
Inserting $d_\text{LW}$ and $d_\text{tid}$ into $R_1$ and $R_2$ 
in eq. (\ref{eq_t_inf}),  the infalling time is calculated 
for the given redshift $z$ and source halo mass $M_\text{source}$.

\end{document}